\documentclass[a4paper,11pt]{article}
\pdfoutput=1 

\usepackage{jheppub}
\usepackage{epsfig}
\usepackage{amsmath}

\def\A0#1{\Pi_{\rm #1}(0)}
\def\AP0#1{\Pi'_{\rm #1}(0)}

\def\be{\begin{equation}}
\def\ee{\end{equation}}
\def\bea{\begin{array}}
\def\eea{\end{array}}
\def\beqa{\begin{eqnarray}}
\def\eeqa{\end{eqnarray}}
\def\beqas{\begin{eqnarray*}}
\def\eeqas{\end{eqnarray*}}

\def\bp{\begin{picture}}
\def\ep{\end{picture}}
\def\bc{\begin{center}}
\def\ec{\end{center}}
\def\bfig{\begin{figure}}
\def\efig{\end{figure}}

\def\bit{\begin{itemize}}
\def\eit{\end{itemize}}
\def\nn{\nonumber}
\def\f{\frac}

\def\[{\left[}
\def\]{\right]}
\def\({\left(}
\def\){\right)}

\def\..{\left.}
\def\.{\right.}
\def\tl{\tilde}
\def\ra{\rightarrow}

\def\da{\dagger}

\def\al{\alpha}
\def\bt{\beta}

\def\ep{\epsilon}

\def\Ga{\Gamma}
\def\ga{\gamma}
\def\pa{\partial}
\def\pr{\prime}

\title{Modulus stabilization of modular flavor models in Jordan frame supergravity}
\author[a]{Fei Wang, \footnote{Corresponding author}}
\author[b]{Ying Kai Zhang}
\affiliation[a]{School of Physics, Zhengzhou University, Zhengzhou 450000, P. R. China}
\emailAdd{feiwang@zzu.edu.cn,ykzhang@zzu.edu.cn}
\abstract{ We propose to discuss the modular flavor model and the stabilization of single modulus field in the Jordan frame supergravity with non-minimal scalar-curvature coupling of the form $\Phi(\tau,\bar{\tau})R$. Modular invariance, positivity of the scale factor and positive definiteness of the Kahler metric constrain stringently the form of the frame function, consequently the Kahler potential by the relation $\Phi(\tau,\bar{\tau})=-3\exp[-K(\tau,\bar{\tau})/3]$. We discuss some general properties of scalar potentials after the scale transformation from the Jordan frame to the Einstein frame.  We find that the shape of the resulting scalar potential in the Einstein frame is quite different from that of ordinary single modulus stabilization mechanism.  The scalar potential could be stationary at the $i\infty$ fixed point, leading to a runaway type vacuum. Such a runaway-type vacuum can be properly stabilized at typical modulus VEV with large $\Im\tau$. We also discuss numerically the modulus stabilization for some simplified scenarios.   }
\begin{document}
\maketitle
\indent
\newpage
\section{\label{sec-1}Introduction}

Torus and toroidal orbifold compactifications, which are frequently employed to derive low-energy four-dimensional effective theories, inherently involve modular symmetry. Geometrically, this symmetry originates from the two-dimensional torus characterized by two independent one-cycles; the freedom to perform basis changes on these cycles corresponds precisely to the ${\rm SL}(2,\mathbb{Z})$ modular transformation. Recently, the application of modular symmetry has achieved significant success in explaining the observed flavor patterns~\cite{Feruglio:2017spp}. In this approach, Yukawa couplings are promoted to modular forms—holomorphic functions transforming properly under the modular group $\Gamma = {\rm SL}(2,\mathbb{Z})$—with the complex modulus field $\tau$ serving as the order parameter for modular flavor symmetry breaking.
A crucial advantage for modular flavor model is that introducing traditional flavon fields is no longer necessary, and all higher-dimensional operators in the superpotential can be unambiguously determined in the limit of unbroken supersymmetry. Since Yukawa couplings are dictated by modular forms, the flavor structures can be highly predictive, requiring very few input parameters once a proper value for the modulus field is determined (for reviews, see~\cite{Feruglio:2019ybq,Kobayashi:2023zzc,Ding:2023htn,Ding:2024ozt}). Furthermore, finite modular flavor symmetries, such as $\Gamma_2\simeq S_3$~\cite{Kobayashi:2018vbk,Okada:2019xqk,Du:2020ylx}, $\Gamma_3\simeq A_4$~\cite{Kobayashi:2018vbk,Petcov:2018snn,Criado:2018thu,Kobayashi:2018scp,Okada:2018yrn,Novichkov:2018yse,Okada:2020ukr,Petcov:2022fjf,CentellesChulia:2023osj,Kumar:2023moh,Nomura:2024ctl,Pathak:2024sei}, $\Gamma_4\simeq S_4$~\cite{Penedo:2018nmg,Petcov:2018snn,Novichkov:2018ovf,Kobayashi:2019xvz,Wang:2020dbp,Qu:2021jdy,deMedeirosVarzielas:2025byb}, and $\Gamma_5\simeq A_5$~\cite{Petcov:2018snn,Novichkov:2018nkm,Ding:2019xna,Yao:2020zml,deMedeirosVarzielas:2022ihu,Abbas:2024bbv}, can be successfully combined with SU(5)~\cite{deAnda:2018ecu,Kobayashi:2019rzp,Du:2020ylx,Chen:2021zty,Zhao:2021jxg,King:2021fhl,Ding:2021zbg}, flipped SU(5)~\cite{Charalampous:2021gmf,Du:2022lij,King:2024gsd}, and SO(10)~\cite{Ding:2021eva,Ding:2022bzs} GUT models, thereby further reducing the number of free parameters.

Ultimately, the optimal value of the modulus field responsible for the observed flavor structure must be derived dynamically through a proper modulus stabilization mechanism. From a top-down perspective, the Vacuum Expectation Value (VEV) of the modulus—which can act as the sole source of flavor symmetry breaking—is determined by the minimum of the modular-invariant scalar potential within 4D $\mathcal{N}=1$ supergravity. This stabilization is also crucial for imparting a non-zero mass to the modulus field, rendering the theory phenomenologically viable. The stabilization of the modulus field within the 4D low-energy effective theories of superstrings and modular flavor models has been extensively discussed in~\cite{Kobayashi:2020hoc,Kobayashi:2020uaj,Ishiguro:2020tmo,NPP:2201.02020,King:2023snq,King:2024ssx,Ding:2024neh,Higaki:2024pql,Higaki:2024jdk,Kobayashi:2023spx,Funakoshi:2024yxg,natural:mu}, primarily utilizing a modular-invariant superpotential originally proposed in~\cite{target:space} (or utilizing specific modular forms coupled to certain matter fields~\cite{Kobayashi:2019xvz,PLNSR:2304.14437}). Once the modular flavor symmetry is fully or partially broken by the VEV of $\tau$, the Yukawa couplings and fermion mass matrices are rigidly fixed. Notably, the desired hierarchical fermion mass matrices can arise naturally if the stabilized modulus lies in close proximity to a fixed point possessing residual symmetry.

In certain well-motivated scenarios, it is highly desirable to stabilize the modulus field near the infinite fixed point $\tau=i\infty$, which preserves a residual $Z^T_N$ symmetry in theories based on modular $\Gamma_N$ invariance. For instance, in finite modular symmetries, stabilizing the modulus at $\Im\tau\gg 1$ ensures that the shift symmetry $\tau\to \tau+1$ remains approximately unbroken. This leads to a residual $Z_N$ symmetry that can be adopted in the Froggatt-Nielsen mechanism, yielding a hierarchy parameter $\exp[-2\pi\Im\tau/N]\ll 1$~\cite{Higaki:2024ueb}. Moreover, it was argued in~\cite{Higaki:2024jdk} that $\Im\tau \gtrsim \mathcal{O}(10)$ is required to generate a highly flat direction in the potential, which can be identified with the axion. Additionally, a modulus VEV with $\Im\tau\gg 1$ provides a natural solution to the $\mu$-problem in the MSSM~\cite{natural:mu}. However, a significant theoretical challenge arises: for most parameter choices in a standard modular-invariant superpotential, the scalar potential of the modulus field diverges as $\tau\to i\infty$. While it is technically possible to stabilize $\Im\tau\gg 1$, existing approaches consistently necessitate the introduction of additional matter fields. Therefore, constructing a simple, minimalist realization that stabilizes $\tau$ near $i\infty$ using solely the modulus field remains a highly desirable and open objective.

Studies of moduli stabilization in heterotic string theory~\cite{target:space,Leedom:2022zdm,Gonzalo:2018guu} indicate that the scalar potential typically favors minima located either at the boundary of the fundamental domain or exclusively along the imaginary axis $\Re\tau=0$ (though exceptions do exist). However, in minimal modular flavor models devoid of flavons, the VEV of $\tau$ can serve as the sole source of CP symmetry breaking~\cite{Baur:2019kwi,Novichkov:2019sqv,Ishiguro:2020nuf,Feruglio:2021dte}. Consequently, it is highly desirable to stabilize $\tau$ strictly inside the fundamental domain of the modular group (rather than on its boundary) to properly act as this CP-breaking source. Nevertheless, detailed analyses reveal that obtaining a viable modulus VEV inside the fundamental domain proves challenging in single-modulus scenarios, unless specific parameter configurations—such as $m\neq 0$ and $n=0$ for the function $H(\tau)$—are chosen. Furthermore, constructing a de Sitter (dS) vacuum with a remarkably tiny positive vacuum energy typically necessitates additional ''matter-like'' fields to provide uplifting contributions. This inevitably complicates the multidimensional search for the true minima of the scalar potential.

In this work, we propose a novel moduli stabilization mechanism that addresses these challenges by introducing non-minimal couplings between the moduli fields and the spacetime curvature tensor. From the perspective of the effective field theory of quantum gravity, a non-minimal coupling between a scalar field $\phi$ and gravity, typically taking the form $|\phi|^2 R$ (where $R$ is the Ricci scalar curvature), naturally and inevitably emerges from graviton-scalar loop corrections. While such couplings have been extensively studied within the contexts of scalar-tensor theories and inflationary cosmology~\cite{Futamase:1987ua,Salopek:1988qh,Makino:1991sg,Fakir:1992cg}, they remain largely unexplored within the modular flavor framework. Specifically, this curvature coupling inherently modifies the scalar potential of the modulus field, possessing the capacity to dynamically generate new minima or reshape existing ones. An analogue of such potential modification is Higgs inflation~\cite{Higgsinflation:A,Bezrukov:2008ut,GarciaBellido:2008ab,DeSimone:2008ei,Bezrukov:2009db,Barvinsky:2009fy}, where an exponentially flat potential at large field values is generated precisely by introducing a non-minimal gravity coupling for the Standard Model Higgs doublet. Therefore, it is compelling to investigate whether this amended scalar potential can successfully stabilize the modulus field near $i\infty$ yet strictly inside the fundamental domain. Such a stabilization would allow the modulus to simultaneously trigger flavor symmetry breaking and naturally generate the hierarchical flavor structures observed in the matter sector. Moreover, it is worthwhile to explore the feasibility of achieving a characteristic CP-violating VEV driven by this modified potential.

This paper is organized as follows: In Section~\ref{sec-2}, we establish the theoretical framework by rigorously defining the modulus supergravity in the Jordan frame, subsequently deriving the modified modular-invariant K\"ahler potential in the physical Einstein frame. In Section~\ref{sec-3}, we present a comprehensive analytical study of the reshaped scalar potential and elaborate on the modulus stabilization mechanism. In Section~\ref{sec-4}, we perform numerical investigations on modulus stabilization across several simplified, benchmark scenarios. Finally, conclusions and outlooks are drawn in Section~\ref{sec-5}. Ultimately, our results suggest that curvature-modulus interactions offer a natural and elegant resolution to some moduli stabilization problems, paving the way for broader applications in modular flavor phenomenology.

\section{\label{sec-2}Modular flavor model in the Jordan frame supergravity}
In the SUSY version of non-minimal coupling of scalar to gravity, the scalar-gravity action in the Jordan frame needs to be supersymmetrized~\cite{Einhorn:2009bh,Ferrara:2010yw,Lee:2010hj}. The Jordan-frame formulation is naturally connected with the superconformal construction of $\mathcal{N}=1$ supergravity. We know that the superconformal theory has a set of local symmetries which includes all $\mathcal{N}=1$ supergravity symmetries and a set of extra local symmetries: local dilatation, $U(1)_R$ symmetry and special supersymmetry. It treats the gravitational coupling $\kappa^{-2}$ as some function of a set of scalars $-\f{1}{3}N(X_I,\bar{X}_I)$. The extra gauge symmetries of the superconformal theory,
including a local conformal symmetry, allow a possibility to derive the supergravity action either in the Einstein frame or in an arbitrary Jordan frame after different gauge fixing for the extra symmetries. In fact, the Einstein frame corresponds to the gauge fixing choices of D-gauge $N(X_I,\bar{X}_I)=-3 M_{P}^2$ while general Jordan frame corresponds to the choice of D-gauge $N(X_I,\bar{X}_I)=\Phi(z_I,\bar z_I)$ with $U(1)$ gauge $y=\bar y$ for variables from the basis $X_I$ to a basis $(y; z_I)$ that satisfies $X_I=y z_I$~\cite{Ferrara:2010yw}.
In this language, the frame function $\Phi(z_I,\bar z_I)$ appears directly as the coefficient of the Ricci scalar in the Jordan frame. Consequently, specifying $\Phi(z_I,\bar z_I)$ provides a geometrically transparent way of encoding the non-minimal couplings and  the symmetry structure of the scalar sector. In particular, the origin of the canonical kinetic terms for all scalars of the MSSM/NMSSM matter contents in the Jordan frame is explained, whereas in the Einstein frame scalar kinetic terms are generally very complicated. Besides, it was noted in~\cite{Kallosh:2000ve} that, there is a simple way to promote any scale-free global SUSY theory to supergravity in the Jordan frame.

 General recipe for the formulation of 4D Jordan frame supergravity was firstly discussed in~\cite{Ferrara:2010yw,Kallosh:2000ve}. As noted previously, the form of Poincare supergravity can be obtained from the underlying superconformal theory, which in addition to local supersymmetry of a Poincare supergravity, has certain extra local symmetries: Weyl symmetry, $U(1)_R$ symmetry, special conformal symmetry and special supersymmetry. General theory of supergravity in an arbitrary Jordan frame was derived in~\cite{Kallosh:2000ve} from the $SU(2,2|1)$ superconformal theory after gauge fixing. 
The  scalar-gravity part of the ${\cal N}=1$, $d=4$ supergravity in a generic Jordan frame with frame function $\Phi(z, \bar z)$, frame function independent Kahler function $K(z,\bar{z})$, and superpotential $W(z)$ is given by~\cite{Ferrara:2010in}
\begin{eqnarray}
\mathcal{L}_{J} = \sqrt{-{g}_{J}}\left[\Phi \left(  -\frac{1}{6} {R}({g}_J)+ {\cal A}^2_\mu(z, \bar z) \right )
+\left( \frac{1}{3}\Phi g_{\alpha\bar\beta }-\frac{\Phi
_\alpha \Phi _{\bar\beta }}{\Phi }\right) \hat {\partial}_\mu z^\alpha \hat {\partial}^\mu \bar z^{\bar\beta }
-V_J \right],
  \label{Jordan}
\end{eqnarray}
with
\begin{eqnarray}
&& \Phi_\alpha \equiv {\frac{\partial }{\partial z^\alpha }}\Phi(z, \bar z),\qquad  \Phi_{\bar\beta } \equiv {\partial\over \partial{\bar z}^{\bar\beta }}\Phi(z, \bar z), \quad  g_{\alpha \bar \beta}= {\frac{\partial^2  {K}(z, \bar z)}{\partial z^\alpha  \partial \bar z^{\bar \beta}}}\equiv {K}_{\alpha \bar \beta} (z, \bar z),
\end{eqnarray}
and ${\cal A}_\mu$ the purely bosonic part of the on-shell value of the auxiliary field $A_\mu$. We adopt $M_P=1$ for the Planck scale in our subsequent discussions (while $M_P$ is kept in some formulas to illustrate the suppression factor).
The Jordan frame potential 
\beqa
V_{J}=\frac{\Phi ^{2}}{9}V_{E}~,  
\label{potential:JordanvsEinstein}
\eeqa
is defined via the Einstein frame potential
\beqa
V_{E}=V_{E}^F +V_{E}^D = e^{\f{K}{M_P^2}}\left( -3\f{W\overline{W}}{M_P^2}
+\nabla_\alpha W g^{\alpha \bar{\beta} }\nabla_{\bar{\beta}}\overline{W}\right) +  \f{1}{2}(\Re
f)^{-1\,AB}P_AP_B\,, \label{VE}
\eeqa
where $\nabla _\alpha W$ denotes the Kahler-covariant
derivative of the superpotential 
\beqa
\nabla_\al W=\pa_\al W+\f{K_\al}{M_P}W~,
\eeqa
and $P_A$ is a momentum map.
A special important class of the superconformal models with
\begin{equation}
 \Phi(z,{\bar z}) = -3 M_P^2 e^{-\f{1}{3M_P^2} {K}(z,{\bar z})}~, \label{framefunction}
\end{equation}
and the  corresponding actions in the Jordan frame
were derived in components in~\cite{Cremmer:1978hn,BFNS-82,CFGVP-1}, and in
superspace in~\cite{Girardi:1984eq,Wess:1992cp}.
In this case the  simpler form of $\mathcal{L}
_{J} $ given by (\ref{Jordan}) was found to be~\cite{Ferrara:2010yw}:
\begin{equation}
\mathcal{L}_{J}=\sqrt{-g_{J}}\left[ \Phi\left( -
\frac{1}{6} R(g_{J})+
{\cal A}^2_\mu(z, \bar z)  \right)
-\Phi _{\alpha\bar\beta }\hat{\partial}
_\mu z^\alpha\hat{\partial}^\mu \bar z^{\bar\beta }-\frac{\Phi^2}{9}V_{E}\right] \, .  
\label{action:Jordan}
\end{equation}
The Kahler metric can be expressed in terms of derivatives with respect to the frame function $\Phi(\tau,\bar{\tau})$
\beqa
K_{\al\bar{\bt}}=-3\f{\Phi_{\al\bar{\bt}}}{\Phi}+3\f{\Phi_{\al}\Phi_{\bar{\bt}}}{\Phi^2}~.
\label{Kahler:metric}
\eeqa

The metric in a Jordan frame is related to the metric in the Einstein frame as follows
\beqa
g^{E}_{\mu\nu}=\Omega^2 g^J_{\mu\nu}~,~~~\Omega^2=-\f{1}{3}\Phi_{total}>0~~.
\label{metric:constraints}
\eeqa
The positivity of scale factor $\Omega^2$ constrains $\Phi(\tau,\bar{\tau})$ to be real negative. 


\subsection{\label{sec-2:1}Modular invariant frame function}
The frame function $\Phi(T)$ for modulus field $T$ needs to be modular invariant so as that its couplings to the Ricci scalar $\Phi(T) R$ remains modular invariant. 
In string theory convention, the real part of modulus field $T$ transforms as scalar and has often the interpretation of volume while the imaginary part is sometimes referred to as T–axion. In the subsequent sections, we adopt the convention in modulus flavor symmetry studies with $\tau \equiv i T$ for modulus field, whose imaginary part transforms as a real scalar and its real part as a pseudoscalar. 

 Ordinary form of the Kahler potential for modular flavor models in rigid SUSY, after proper Kahler transformation, can be promoted to the frame function by the relation~(\ref{framefunction}).
Since 
\beqa
(\tau-\bar{\tau})\ra  |c\tau+d|^{-2}(\tau-\bar{\tau})~,
\eeqa
under modular transformation, the combination
\beqa
G_0(\tau,\bar{\tau})\equiv -i(\tau-\bar{\tau})\eta^2(\tau)\overline{\eta^2(\tau)}~,
\eeqa
is modular invariant. In general, the real part of Petersson norm type modular invariants given by
\beqa
\Re[(\Im(\tau))^{2k} f(\tau)\overline{g({\tau})}]~,
\eeqa
with $f(\tau)$, $g(\tau)$ modular forms of weight $k$,  can also be adopted. On the other hand, which will become clear later, negativity of the modular invariants frame function as well as the positive definiteness of the corresponding Kahler metric set very stringent constraints on the choices of such modular invariants. For example, $G\equiv [\Im(\tau)]^{2k} \left|f(\tau)\right|^2$, which can act as a multiple factor for $\Phi(\tau,\bar{\tau})$, is positive on the upper half plane and will not always give negative definite $G_{\tau,\bar{\tau}}$. Therefore, it will not always give positive definite contributions to the total Kahler metric $K_{\tau\bar{\tau}}$ according to eq.(\ref{Kahler:metric}). 

It is known that any $SL(2,Z)$-invariant meromorphic function can be expressed as a function of Klein $j$-function.
Therefore, for simply (and also to give properly the ordinary no-scale type Kahler potential for modulus field), we can parameterize the general modular invariant form of frame function as
\beqa
\tl{F}[G_0(\tau,\bar{\tau}),H(\tau),\overline{H(\tau)}],
\eeqa
which denotes arbitrary real valued function of $G_0(\tau,\bar{\tau})$, $H(\tau)$ and $\overline{H(\tau)}$. Here the modular invariant holomorphic function $H(\tau)$ is given by
\beqa
{H}(\tau)&=&\(j(\tau)-1728\)^{\f{m}{2}}j(\tau)^{\f{n}{3}}{{\cal P}}\(j(\tau)\),\nn\\
 &\equiv&\(\f{G_6(\tau)}{\[\eta(\tau)\]^{12}}\)^{{m}}\(\f{G_4(\tau)}{\[\eta(\tau)\]^{12}}\)^{{n}}
 {\cal P}\(j(\tau)\)~,
 \label{modular:invariant}
\eeqa
where ${\cal P}\(j(\tau)\)$ denotes a polynomial with respect to  Klein $j$-function $j(\tau)$ and $m$,$n$ are non-negative integers. It is obvious from the expression that non-trivial $H(\tau)$ diverges at the value $\tau=i\infty$.

 The modular invariant form of frame function $\Phi(\tau,\bar{\tau})$ for modulus field can be chosen to be
\beqa
\Phi(\tau,\bar{\tau})&=&-3\[-i(\tau-\bar{\tau})\eta^2(\tau)\overline{\eta^2(\tau)}\]\[\overline{{H}(\tau)}{H}(\tau)\]
\exp\left\{\xi\[H(\tau)+\overline{H(\tau)}\]\right\},
\eeqa
and the complete frame function is given by
\beqa
\Phi_{total}&=&\Phi(\tau,\bar{\tau})+\Phi_{matter}(\phi^{(I_n)},\bar{\phi}^{(I_n)})~,
\eeqa
with $\Phi_{matter}$ the modular invariant frame function for MSSM/NMSSM matter fields $\phi^{(I_n)}$ of modular weight $k_{I_n}$ (including the Higgs superfields), which is given by
\beqa
\Phi_{matter}(\phi^{(I_n)},\bar{\phi}^{(I_n)})=\sum_n \f{\bar{\phi}^{(I_n)}\phi^{(I_n)}}{[-i(\tau-\bar{\tau})]^{k_{I_n}}}~.
\eeqa 
 It can be seem from eq.(\ref{action:Jordan}) that, the prefactor of the kinetic term for $\tau$ in Jordan frame is given by
\beqa
{\Phi_{\tau\bar{\tau}}}&=&e^{-\f{1}{3}K(\tau,\bar{\tau})}\[K_{\tau\bar{\tau}}-\f{1}{3} |K_\tau|^2\].
\eeqa
The matter fields can have canonical kinetic terms in the Jordan frame up to modular weight related constant factor after the modulus field is fixed. Note that the modular invariant contributions of matter fields to complete frame function is assumed to take the canonical form. The most general choices of matter frame function consistent with the symmetries of the model contains additional terms with additional parameters, which reduce the predictive power of these constructions~\cite{Chen:2019ewa}. To solve the unconstrained frame function (Kahler potential) problem, eclectic flavor group scheme can be adopted~\cite{Baur:2019kwi}.

We should note again that there are other modular invariant combinations that can be adopted in the frame function. For example, the non-holomoprhic Eisenstein serie, with $y=\Im\tau$, given by
\beqa
E(\tau,s)=\f{1}{2}\sum_{\substack{(m,n)\in Z^2;\\gcd(m,n)=1}}\f{y^s}{|m\tau+n|^{2s}}~,~~~{\Re s}>1
\eeqa
can be introduced in the frame function $\Phi\propto -3[-i(\tau-\bar{\tau})\eta^2(\tau)\overline{\eta^2(\tau)}]e^{-E(\tau,s)}$. It is a Maass form that satisfies the Laplacian equation
\beqa
\Delta_L E(\tau,s)\equiv y^2 \pa_\tau\pa_{\bar{\tau}}E(\tau,s)=s(s-1)E(\tau,s)~.
\eeqa
It can be checked that the negativity of the frame function $\Phi$ as well as the positive definiteness of the corresponding Kahler metric for real $s>1$ can be satisfied. We will discuss such possibilities in our subsequent works for mulitple-modulus cases~\cite{FW-YKZ}.


From the relation between the frame function and the Kahler potential
\beqa
\Phi_{total}=-3 M_P^2 e^{-\f{1}{3 M_P^2}K_{total}}~,
\eeqa
we can obtain the $N=1$ supergravity action in the Jordan frame, with the modular invariant Kahler potential given by
\beqa
K_{total}&=&-3\ln\left\{-i(\tau-\bar{\tau})\left|\eta(\tau)\right|^4\left|{H}(\tau)\right|^2
\exp\left\{\xi\[H(\tau)+\overline{H(\tau)}\]\right\}-\f{\Phi_{matter}(\phi^{(I_n)},\bar{\phi}^{(I_n)})}{3M_P^2}\right\}.\nn\\
\eeqa
In the small matter field limits, the Kahler potential can be approximately given as
\beqa
K_{total}&\approx&-3\left\{\ln\[-{i}(\tau-\bar{\tau})\left|\eta(\tau)\right|^4\]+\ln\[\overline{H(\tau)}H(\tau)\]\right.\nn\\
&+&\left.\xi\[H(\tau)+\overline{H(\tau)}\]\right\}-\f{3}{\Phi(\tau,\bar{\tau})}\sum\limits_n \f{\bar{\phi}^{(I_n)}\phi^{(I_n)}}{M_P^2 [-i(\tau-\bar{\tau})]^{k_{I_n}}},
\label{Ktotal}
\eeqa
with non-canonical kinetic terms for matter fields. Note that the small field limits always hold for MSSM matter contents, as the quantum fluctuations from the vanishing or electroweak scale VEVs are always small in comparison to the Planck scale. The positive definite Kahler metric $K_{\tau\bar{\tau}}={3}/{4[\Im{\tau}]^2}$ adopted in ordinary modulus stabilization mechanism discussions is kept here in the small $\phi^{(I_n)}$ limit for matter fields.

In the Jordan frame, the superpotential $W(\Phi)$ should be modular invariant because of the modular invariance of frame function (consequently, the modular invariance of the Kahler potential by the relation (\ref{framefunction})) and the combination $e^K|W|^2$. 
The modular invariant superpotential can take the form
\beqa
W&=&\Lambda^3 \tl{H}(\tau)=\tilde{c}_0^3 M_{Pl}^3 \(j(\tau)-1728\)^{\tl{m}/2}j(\tau)^{\tl{n}/3}{\tl{\cal P}}\(j(\tau)\),
\label{superpotential}
\eeqa
with the most general form of $\tl{\cal P}(j(\tau))$ given by
\beqa
\tl{\cal P}(j(\tau))=\sum\limits_{k=0}^{\tl{N}}  c_k\[j(\tau)\]^k~,
\eeqa 
Here $\Lambda\equiv \tilde{c}_0 M_{pl}$ with real $\tilde{c}_0$. In our subsequent studies, we choose $\tilde{c}_0=10^{-3}$ so as that $\Lambda$ is typically of order the GUT scale. The modular invariant holomorphic function $\tl{H}(\tau)$ takes the same general form as $H(\tau)$ with independent choice of polynomial $\tl{P}(j(\tau))$ on $j(\tau)$ and the replacement of integers $m\ra\tl{m}$,  $n\ra\tl{n}$.

From eq.(\ref{potential:JordanvsEinstein}), the supergravity scalar potential in the Jordan frame $V_J$ can be obtained by the scalar potential $V_E$ in the Einstein frame 
\beqa
V_E&=&e^{K}\[K_{\tau\bar{\tau}}^{-1}\left|\pa_\tau W+{K_\tau} W\right|^2+
\sum\limits_{I=\tau,\phi^{(I_n)}}K_{I\bar{J}}^{-1}\left|\pa_I W+{K_I} W\right|^2 -3 {\left|W\right|^2}\]~,\\
&=&\f{27\Lambda^6\[K_{\tau\bar{\tau}}^{-1} \left|\tl{H}^\pr(\tau)+C_1(\tau,\bar{\tau})
\tl{H}(\tau) \right|^2-3 \left|\tl{H}(\tau)\right|^2\]}{\[-i(\tau-\bar{\tau})\eta^2(\tau)\overline{\eta^2(\tau)}|{H}(\tau)|^2
e^{\xi\[H(\tau)+\overline{H(\tau)}\]}-\f{1}{3}\sum\limits_n \f{\bar{\phi}^{(I_n)}\phi^{(I_n)}}{[-i(\tau-\bar{\tau})]^{k_{I_n}}}\]^3}+V(\phi^{(I_n)})~,\nn
\eeqa
with $V(\phi^{(I_n)})$ the terms containing $\phi^{(I_n)}$  that proportional to  $K_{\phi^{(I_n)},\bar{\phi}^{(I_n)}}^{-1}, K_{\phi^{(I_n)},\bar{\tau}}^{-1}$ etc. As we are interested in the stabilization of modulus field, which can safely take the small matter field limits, the $\phi^{(I_n)}$ relevant terms can be neglected in our subsequent discussions. 
Within the Einstein frame scalar potential
\beqa
K_\tau&\equiv&C_1(\tau,\bar{\tau})\\
&\simeq &-\f{3\[i\eta^2(\tau)\overline{\eta^2(\tau)}
+2i(\tau-\bar{\tau})\eta(\tau)\eta^\pr(\tau)\overline{\eta^2(\tau)}\]}{\[i(\tau-\bar{\tau})\eta^2(\tau)\overline{\eta^2(\tau)}\]}-3\f{{H}^\pr(\tau)}{{H}(\tau)}
-3\xi{H}^\pr(\tau)~\nn\\
&=&\f{3}{2\Im\tau}-\f{3i}{2\pi}G_2(\tau)-3\f{{H}^\pr(\tau)}{{H}(\tau)}-3\xi{H}^\pr(\tau)~,
\eeqa
and $K_{\tau\bar{\tau}}^{-1}=1/K_{\tau\bar{\tau}}$ is given by
\beqa
K_{\tau\bar{\tau}}&=&\f{3}{\[-i(\tau-\bar{\tau})\]^2}.
\eeqa

The invariance of the superpotential $W(\Phi)$ under the modular flavor group requires that the expansion of $W(\Phi)$ in power series of the supermultiplets $\phi^{(I)}$:
\beqa
W_{J}(\phi)=\sum_n Y_{I_1...I_n}(\tau)~ \phi^{(I_1)}... \phi^{(I_n)}~~~.
\label{Jordan:superpotential}
\eeqa
with $Y^{k_Y(n)}_{I_1...I_n}(\tau)$ the modular forms of weight $k_Y(n)$ transforming
in the representation $\rho$ of $\Gamma_N$:
\beqa
Y^{k_Y(n)}_{I_1...I_n}(\gamma\tau)=(c\tau+d)^{k_Y(n)} \rho(\gamma)~Y_{I_1...I_n}(\tau)~~~,
\eeqa 
and
\beqa
\rho\otimes\rho^{I_1}\otimes...\otimes\rho^{I_n}\supset {\bf 1}~,~~~~~~~\sum\limits_{a=1}^{n} k_{I_a}=k_Y(n)~.
\eeqa

After the scaling with $g_{\mu\nu}^E=\Omega^2 g_{\mu\nu}^J$, the curvature changes as
\beqa
R_J=\Omega^{2}\[R_E+6\square_E\ln\Omega-6g^{E,\mu\nu}(\nabla_\mu\ln\Omega)(\nabla_\nu\ln\Omega)\]~,
\eeqa
with
\beqa
\square_E\ln\Omega=g_E^{\mu\nu}\nabla_\mu\nabla_\nu\ln\Omega=\f{1}{\sqrt{-g_E}}\pa_\mu\(\sqrt{-g_E}g_E^{\mu\nu}\pa_\nu\ln\Omega\)~.
\eeqa
The kinetic term of $\tau$, in addition to contributions from ${\cal A}^2_\mu$, is no-longer canonical with
\beqa
\f{{\cal L}_E}{\sqrt{-g_E}}&\supseteq& \f{M_P}{2} R_E+K_{\tau\bar{\tau}}g_E^{\mu\nu}\pa_\mu \tau\pa_\nu\bar{\tau}+\f{1}{[-i(\tau-\bar{\tau})]^{k_{\Psi}}}i\f{\bar{\Psi}_A}{\Omega^{\f{3}{2}}}\ga^a e_a^\mu(D_\mu+\f{1}{8}\omega_{\mu ab}[\ga^{a},\ga^b])\f{\Psi_A}{\Omega^{\f{3}{2}}}\nn\\
&+&Z_{\overline{\phi^{(I_n)}}\phi^{(I_n)}}\[(D_\mu \phi^{(I_n)})^\da (D^\mu \phi^{(I_n)})\]-\f{1}{\Omega^4}{Y_{{\bf r};AB}^{(k)}(\tau)}\bar{\Psi}_A H \Psi_B-\f{1}{\Omega^4}V_{scalar}+\cdots~.\nn\\ 
\label{action}
\eeqa
Note that the Kahler metric for matter fields $\phi^{(I_n)}$ 
\beqa
K_{\phi^{(I_m)}\overline{\phi^{(I_n)}}}&=&
-3\f{\Phi_{total}\Phi_{total;\phi^{(I_m)}\overline{\phi^{(I_n)}}}-\Phi_{total;\phi^{(I_m)}}\Phi_{total;\overline{\phi^{(I_n)}}}}{\Phi_{total}^{2}}\nn\\
&=&-\f{3}{[-i(\tau-\bar{\tau})]^{k_{I_n}}}\f{\Phi_{total}\delta_{mn}
-\f{\overline{\phi^{(I_m)}}\phi^{(I_n)}}{[-i(\tau-\bar{\tau})]^{k_{I_m}}}}{\Phi_{total}^2}~,
\eeqa
 can lead to the prefactor for the scalar kinetic term 
\beqa
Z_{\overline{\phi^{(I_n)}}\phi^{(I_n)}}&=&\f{1}{\Omega^2[-i(\tau-\bar{\tau})]^{k_{I_n}}}
+\f{\overline{\phi^{(I_n)}}\phi^{(I_n)}}{3\Omega^4[-i(\tau-\bar{\tau})]^{2k_{I_n}}}~,
\eeqa
with the first part from Weyl scaling of $\sqrt{-g_J}g^{\mu\nu}_J$ in the scalar kinetic term and the second part from the transformation of the Ricci curvature $R_J$.
The vierbein ${e}_{\mu}^{\alpha}$ in (\ref{action}) represents the square root of the metric, which satisfies 
\beqa {e}_\mu^{a}{e}_\nu^{b}\eta_{ab}&\equiv& g_{\mu\nu}~,~~~~~~{e}^{a\mu}\equiv e^a_\rho g^{\rho\mu}~\nn\\
      {e}_\mu^{a} e^\mu_b&=&\delta^a_b~,~~~~~~~{e}_\nu^{a} e^\mu_a=\delta^\mu_\nu~.
\eeqa
 We can rewrite $\sqrt{-g}=|{e}_\mu^{a}|\equiv {e}$ and $\gamma^\mu={e}^{\mu}_{a}\gamma^{a}$. The spin connection can be expressed in terms of vierbein as
\beqa
\omega_\mu^{ab}
&=&\f{1}{2}e^{\nu a}\(\pa_\mu e_\nu^b-\pa_\nu e^b_\mu\)-\f{1}{2}e^{\nu b}\(\pa_\mu e_\nu^a-\pa_\nu e^a_\mu\)+\f{1}{2}e^{\nu a}e^{\sigma b}(\pa_\sigma e_{\nu c}-\pa_\nu e_{\sigma c}) e^c_\mu~.
\eeqa
The fermionic matter fields scale as $\Omega^{-3/2}$ while the scalar potential terms scale as $\Omega^{-4}$ under metric scaling. The normalization of MSSM/NMSSM matter fields are non-trivial for complex scalars. 
Defining $\phi_{I_n}=r_{I_n} e^{i\theta_{I_n}}$ for complex scalar $\phi_{I_n}$, the kinetic term can be written as
\beqa
&&\[\f{1}{\Omega^2 (2\Im\tau)^{k_{\phi_I}}}+\f{\overline{\phi^{(I_n)}}\phi^{(I_n)}}{3\Omega^4(2\Im\tau)^{2k_{\phi_I}}}\](\pa^\mu\phi_{I_n})^*(\pa_\mu\phi_{I_n})\nn\\
&=&\[\f{1}{\Omega^2 (2\Im\tau)^{k_{\phi_I}}}+\f{r^2_{I_n}}{3\Omega^4(2\Im\tau)^{2k_{\phi_I}}}\]\(\pa_\mu r_{I_n} \pa^\mu r_{I_n}+r_{I_n}^2\pa_\mu\theta_{I_n}\pa^\mu\theta_{I_n}\)~,\nn\\
&\equiv&\(\pa_\mu \tilde{r}_{I_n} \pa^\mu \tilde{r}_{I_n}+\tilde{r}_{I_n}^2\pa_\mu\tilde{\theta}_{I_n}\pa^\mu\tilde{\theta}_{I_n}\)
\eeqa
We can normalize the kinetic term by changing the variables from $(r_{I_n},\theta_{I_n})$ to normalized field 
$(\tilde{r}_{I_n},\tilde{\theta}_{I_n})$ by
\beqa
\f{d \tilde{r}_{I_n}}{d {r}_{I_n}}&=&\sqrt{\f{1}{\Omega^2 (2\Im\tau)^{k_{\phi_I}}}+\f{r^2_{I_n}}{3\Omega^4(2\Im\tau)^{2k_{\phi_I}}}},\nn\\
\f{d \tilde{\theta}_{I_n}}{d{\theta}_{I_n}}&=&\f{{r}_{I_n} }{\tilde{r}_{I_n}({r}_{I_n})}\sqrt{\f{1}{\Omega^2 (2\Im\tau)^{k_{\phi_I}}}+\f{r^2_{I_n}}{3\Omega^4(2\Im\tau)^{2k_{\phi_I}}}}~.
\eeqa 
The first equation admits an exact solution, which can be obtain easily from the similar differential equation in Higgs inflation~\cite{Garcia-Bellido:2008ycs,Rubio:2018ogq}. The second equation can be directly solved by a simple scaling, as the RHS of the differential equation is independent of $\theta_{I_n}$. From the solution, we have 
\beqa
\left\{\bea{c}\tilde{r}_{I_n}=\f{{r}_{I_n}}{\sqrt{-\f{\Phi(\tau,\bar{\tau})}{3}(2\Im\tau)^{k_{\phi_I}}}}~,~ \nn\\
{\tilde{\theta}_{I_n}}={{\theta}_{I_n}}~.~~~~~~~\quad\quad\quad\eea \right. ~~~{\rm for}~~~{r}_{I_n}\ll {-\f{\Phi(\tau,\bar{\tau})}{3}(2\Im\tau)^{k_{I_n}}} M_P,
\eeqa
which amounts to a constant scaling of the scalar field after the modulus field is stabilized.

Such discussions for $\phi_{I_n}$ can be applied to the Higgs fields $H_u$ and $H_d$. Similar discussions can also be given for fermion fields. In fact, supersymmetry ensures that the fermionic component and the scalar component of a chiral superfield transform in collective manner, except that the scaling dimension of fermions (which is $w=3/2$) is different to that of the scalars (which is $w=1$).
Therefore, in small (matter) field regions after normalization, the Yukawa couplings with canonical (up to modular weight related constant factor) matter and Higgs fields in the Jordan frame no longer need any scale factor when the Jordan frame is transformed to Einstein frame in low energy MSSM/NMSSM.

Non-minimal kinetic term for modulus field can be normalized to a canonical one by making the transformation between the non-canonical variable $\tau$ and the canonical ones $\tl{\tau}$, which is represented by $\tl{\tau}=f(\tau)$. The Yukawa couplings for the matter superfields are just modular forms $Y_{\bf r}^{(k)}(\tau)$ in the Jordan frame~\footnote{Note that composing a holomorphic function with a modular form, i.e., taking $g \circ f$ where $g$ is a modular form and $f$ is a holomorphic function, generally cannot be expressed as a combination of modular forms. So, it is not meaningful to assign $Y_{\bf r}^{(k)}(\tl{\tau})\equiv Y_{\bf r}^{(k)}[f^{-1}(\tau)]$ as Yukawa couplings for the realization of modular flavor model in the Jordan frame.}, which take values on the VEV of the non-canonical modulus field $\tau$. As we concentrate only on the moduli stabilization for $\tau$ that appear as the modulus variable in modular forms $Y_{\bf r}^{(k)}(\tau)$, we need only to concentrate on the potential and stabilization for the $\tau$ field instead of the concrete value of canonical field $\tl{\tau}$.

\section{\label{sec-3}Minimum of the scalar potential}
It is known that the VEV of the modulus field $\tau$ generally break the modular symmetry completely, except at three inequivalent fixed points $\tau=i\infty, i, e^{i2\pi/3}$ in the fundamental domain. 
The modular invariant scalar potential is also CP invariant when the coefficients of the polynomials with respect to $j(\tau)$ in $P(j(\tau))$ and $\tl{P}(j(\tau))$ are real.
The reality of the scalar potential, together with the $S$ and $T$ modular transformation, ensures that the first derivative along certain directions at the boundary of the fundamental domain vanishes. 
Hence, the scalar potential is always stationary at the finite fixed points $\tau = \omega = e^{2\pi i/3}$ and $\tau=i$, as they are the intersections of the CP
invariant lines and two independent directional derivatives of the scalar potential vanish there. Note that the CP-conserving VEVs of $\tau$ are the imaginary axis and the boundaries of the fundamental domain\footnote{When non-trivial contributions such as radiative corrections are considered, the VEV may deviate from the fixed points~\cite{NPP:2201.02020,Higaki:2024pql}}.

In principle, to determine the VEV of the modulus field, we need to determine the minimum conditions of scalar potential for all fields, including the modulus field and all the MSSM/NMSSM matter fields. The VEVs of matter fields are mostly vanishing (or at most of order the electroweak scale, much smaller than the typical modulus stabilization scale), so as that terms involving the VEVs for matter fields in the equations of extrema conditions vanish (or can be safely neglected). Therefore, we can safely neglect the matter part of the scalar potential and concentrate only on the pure modulus part of scalar potential for modulus stabilization discussions, which reads
\beqa
V_E&\simeq&e^{K}\[K_{\tau\bar{\tau}}^{-1}\left|\pa_\tau W+{K_\tau} W\right|^2|^2 -3 {\left|W\right|^2}\]~,\nn\\
&=&\f{27\Lambda^6\[K_{\tau\bar{\tau}}^{-1} \left|\tl{H}^\pr(\tau)+C_1(\tau,\bar{\tau})
\tl{H}(\tau) \right|^2-3 \left|\tl{H}(\tau)\right|^2\]}{\[-i(\tau-\bar{\tau})\eta^2(\tau)\overline{\eta^2(\tau)}|{H}(\tau)|^2
e^{\xi\[H(\tau)+\overline{H(\tau)}\]}\]^3}~.
\label{tauscalar:potential}
\eeqa

The minimum conditions of the scalar potential are determined by the derivatives $V^E_\tau$ and $V^E_{\bar{\tau}}$ as
\beqa
\f{\pa }{\pa\tau}V_E(\tau,\bar{\tau})=\f{\pa }{\pa\bar{\tau}}V_E(\tau,\bar{\tau})=0~,\label{V:tau}
\eeqa
which are weights $(2,0)$ and $(0,2)$ non-holomorphic modular functions, respectively. The second one is obtained from the first one by complex conjugation. 
To determine if a local extremum is a minimum, we need to assess the positive definiteness of the $2\times2$ Hessian matrix 
\beqa
(H_f)_{ij}\equiv \f{\pa^2 V}{\pa x_i\pa x_j}~,~~~{\rm for}~~x_i={\Re}\tau, {\Im}\tau,
\eeqa
as well as the positivity of $V_{x_1x_1}$ at that extremum. The elements of the Hessian matrix can be rewritten by
\beqa
V_{x_1x_1}&\equiv& \f{\pa^2 }{\pa x_1^2}V =2\(\f{\pa^2}{\pa\tau\pa\tau^*}+\Re\f{\pa^2 }{\pa \tau\pa\tau}\)V(\tau,{\tau}^*),\nn\\
V_{x_2x_2}&\equiv&\f{\pa^2 }{\pa x_2^2}V =2\(\f{\pa^2}{\pa\tau\pa\tau^*}-\Re\f{\pa^2 }{\pa \tau\pa\tau}\)V(\tau,{\tau}^*),\nn\\
V_{x_1x_2}&\equiv&\f{\pa^2 }{\pa x_1\pa x_2}V =-2\Im \f{\pa^2 }{\pa \tau\pa\tau}V(\tau,{\tau}^*),
\eeqa

Given the form of the frame function, the derivative $V^E_\tau$ can be calculated to be
\beqa
V_{E;\tau}&=&\f{-27\[K_{\tau\bar{\tau}}^{-1} \left|\nabla_\tau W \right|^2-3\Lambda^6 \left|\tl{H}(\tau)\right|^2\]^\pr}
{\Phi^3}+\f{81\[K_{\tau\bar{\tau}}^{-1} \left|\nabla_\tau W \right|^2-3\Lambda^6 \left|\tl{H}(\tau)\right|^2\]\Phi_\tau}{\Phi^4},\nn\\
\label{Fderivative:V}
\eeqa 
with
\beqa
{\Phi_\tau}&=&\[i-\f{G_2(\tau)}{2\pi}(\tau-\bar{\tau})\]{\Phi}
+3\[i(\tau-\bar{\tau})\eta^2(\tau)\overline{\eta^2(\tau)}\]\[\overline{{H}(\tau)}H^\pr(\tau)\]\exp\left\{\xi\[H(\tau)+\overline{H(\tau)}\]\right\}~.\nn\\
&+&3\[i(\tau-\bar{\tau})\eta^2(\tau)\overline{\eta^2(\tau)}\]\[\overline{{H}(\tau)}H(\tau)\]\exp\left\{\xi\[H(\tau)+\overline{H(\tau)}\]\right\}\xi H^\pr(\tau)~,
\eeqa
and 
\beqa
&&\[K_{\tau\bar{\tau}}^{-1} \left|\nabla_\tau W \right|^2-3\Lambda^6 \left|\tl{H}(\tau)\right|^2\]^\pr\\
&=& \(K_{\tau\bar{\tau}}^{-1}\)_{\tau} \nabla_\tau W \overline{\nabla_\tau W}+K_{\tau\bar{\tau}}^{-1} (\nabla_\tau W)^\pr\overline{\nabla_\tau W}+K_{\tau\bar{\tau}}^{-1} (\nabla_\tau W)(\overline{\nabla_\tau W})^\pr
-3\Lambda^6 \tl{H}^\pr(\tau)\overline{\tl{H}(\tau)}~.\nn
\eeqa
Note that
\beqa
(\nabla_\tau W)^\pr
&=&\Lambda^3\[\tl{H}^\pr(\tau)+C_1(\tau,\bar{\tau})
\tl{H}(\tau)\]^\pr~, \\
&=&\Lambda^3\[\tl{H}^{\pr\pr}(\tau)+\[C_1(\tau,\bar{\tau})\]_\tau
\tl{H}(\tau)+C_1(\tau,\bar{\tau})\tl{H}^\pr(\tau)\]~,\nn
\eeqa
and
\beqa
(\nabla_\tau W)(\overline{\nabla_\tau W})^\pr=\Lambda^3\[\tl{H}^\pr(\tau)+C_1(\tau,\bar{\tau})
\tl{H}(\tau)\]\[\overline{C_1(\tau,\bar{\tau})}\]_\tau \overline{\tl{H}(\tau)}~,
\eeqa
both of which can divide $\tl{H}(\tau)$  when $\tl{H}(\tau)\neq 0$ (see Appendix \ref{appendix:B}). Therefore, the combination $|\tl{H}(\tau)|^2$ can be factored out from $V_{E;\tau}$ for $\tl{H}(\tau)\neq 0$.  

It can be checked that $K^{-1}_{\tau\bar{\tau}}$ as well as $(K^{-1}_{\tau\bar{\tau}})_\tau$ are not finite at $\tau=i\infty$. The asymptotic behavior of $K^{-1}_{\tau\bar{\tau}}$ at $\tau=i\infty$ is given by
\beqa
\left.K^{-1}_{\tau\bar{\tau}}\right|_{\tau=i\infty}=\lim_{\Im\tau\ra\infty}\f{4}{3}\(\Im\tau\)^2~, ~
\eeqa
while the asymptotic behavior of $(K^{-1}_{\tau\bar{\tau}})_\tau$ is
\beqa
(K^{-1}_{\tau\bar{\tau}})_\tau=-\f{K_{\tau\bar{\tau}\tau}}{\(K_{\tau\bar{\tau}}\)^2}=-i\lim_{\Im\tau\ra\infty}\f{4}{3}\Im\tau~.
\eeqa

Similarly, it can be checked that both $\[C_1(\tau,\bar{\tau})\]$ and $\[C_1(\tau,\bar{\tau})\]_\tau$ are divergent at $\tau=i\infty$ while $\[\overline{C_1(\tau,\bar{\tau})}\]_\tau$ is finite at $\tau=i\infty$. The asymptotic behavior of them are given by
\beqa
\left.C_1(\tau,\bar{\tau})\right|_{\tau=i\infty}& \sim& -3\tl{A}_H-3\xi \tl{A}_H H(i\infty)~,\nn\\
\left.\[C_1(\tau,\bar{\tau})\]_\tau\right|_{\tau=i\infty}& \sim& -3\tl{B}_H+3\tl{A}_H^2-3\xi \tl{B}_H H(i\infty)~.
\eeqa

 Because of the $1/\Phi^3$, $\Phi_\tau/\Phi^4$ factors in the expression of $V_{E;\tau}$ and the relation
 \beqa
 \exp\[ H(i\infty)\]\gg \[H(i\infty)\]^M,~~{\rm for}~~~H(i\infty)\sim  +\infty~, 
 \eeqa 
for any degree $M$ and non-trivial $H(\tau)$, it can be shown that $\tau=i\infty$ is a solution of $V_{E;\tau}=0$ when $\xi>0$. 
 Besides, from the discussions in Appendix \ref{appendix:B}, it can be checked that $\tau=i\infty$ is also a solution of $V_{E;\tau}=0$ for
\beqa
\tilde{k}\equiv 6\(\f{{m}}{2}+\f{{n}}{3}+{N}\)-\f{1}{2}-2\(\f{\tl{m}}{2}+\f{\tl{n}}{3}+\tl{N}\)\geq 0~,  
\label{minimum:iinfty}
\eeqa
when $\xi=0$, assuming that $P(j(\tau))$ and $\tl{P}(j(\tau))$ are degree $N$ and $\tl{N}$ polynomials in $j(\tau)$. The condition for equality in the inequality (\ref{minimum:iinfty}) can hold because of the $(2\Im\tau)^3$ factor in the denominator of $V_{E;\tau}$. However, $\tau=i\infty$ is not a solution of $V_{E;\tau}=0$ when $\xi<0$.  Therefore, there is a run-away type local minimum at $\tau=i\infty$ for some input parameter choices, where the residual $Z_N$ symmetry is unbroken. 

When $\tau=i\infty$ is a solution of $V_{E;\tau}=0$, the corresponding vacuum energy vanishes for $\xi>0$. When $\xi=0$, the vacuum energy at $\tau=i\infty$ can be calculated to be
\beqa
V_E(i\infty)&=&
\left.\f{\Lambda^6}{\[(2\Im\tau)\eta^2(\tau)\overline{\eta^2(\tau)}\]^3
\left|{H}(\tau)\right|^6}\(-3\tl{A}_H\)\f{4}{3}\(\Im\tau\)^2\left|\tl{H}(\tau)\right|^2\right|_{\tau=i\infty}~,\nn\\
&=&\f{2\Lambda^6}{(3\Im\tau)}\(-3\tl{A}_H\)\tl{q}^{-2\(\f{\tl{m}}{2}+\f{\tl{n}}{3}+\tl{N}\)+6\(\f{{m}}{2}+\f{{n}}{3}+{N}\)-\f{1}{2}}~,
\eeqa
with $\tl{q}=e^{-2\pi \Im\tau}$ for $\Im\tau\ra\infty$.
Obviously, $V_E(i\infty)$ vanished when 
\beqa
6\(\f{{m}}{2}+\f{{n}}{3}+{N}\)-\f{1}{2}-2\(\f{\tl{m}}{2}+\f{\tl{n}}{3}+\tl{N}\)\geq 0~,
\eeqa
agree with the condition (\ref{minimum:iinfty}). 

The run away vacuum at $i\infty$ still needs to be stabilized at some point near $i\infty$ (for example, $\Im\langle\tau\rangle=i\tau_0$ with large real $\tau_0$). The simplest possibility is to introduce soft SUSY breaking scalar masses for modulus field, which takes the form
\beqa
{\cal L}\supset -m_{soft}^2 |\tau|^2~, 
\label{runaway:one}
\eeqa
with $m_{soft}$ typically of order TeV so as that $m^2_{soft}\sim 10^{-15}$ for $M_{Pl}=1$. Therefore, the scalar potential for $\tau$, whose asymptotic behavior near $i\infty$ scales like 
\beqa 
V_{\tau}\sim  \left\{\bea{c}e^{-3\xi\[H(\tau)+\overline{H(\tau)}\]}+m_{soft}^2|\tau|^2\sim e^{-6\xi (e^{-2\pi i\tilde{l}_N\tau})}+m_{3/2}^2|\tau|^2~,~~~~~~~~\xi>0~~~ \\
e^{2\tilde{k}_N\pi i\tau}+m_{soft}^2|\tau|^2~,~~~~~~~~~~~~~~~~~~~~~~~~~~~~~~~~~~~~~~~~~~~~~~~~~~~~~~\xi=0~~~\eea\right.
\eeqa 
with the coefficient
\beqa
\tilde{l}_N&\equiv& \f{\tl{m}}{2}+\f{\tl{n}}{3}+\tl{N}~,\nn\\
\tilde{k}_N&\equiv& 6\(\f{{m}}{2}+\f{{n}}{3}+{N}\)-\f{1}{2}-2\(\f{\tl{m}}{2}+\f{\tl{n}}{3}+\tl{N}\)~.
\eeqa
It can be calculated that, with such soft SUSY breaking contributions, the modulus field is stabilized at
\beqa
\Im\langle\tau\rangle\sim\f{1}{2\tilde{k}_N\pi}W(\f{2\tilde{k}_N^2\pi^2}{m_{soft}^2})\sim \f{10}{\tilde{k}_N}~,
\label{xizerotau:VEV}
\eeqa
when $\xi=0$, with the approximation expression of the Lambert W-function 
\beqa
W(z)\ra \ln z-\ln\ln z+\f{\ln\ln z}{\ln z}~,
\eeqa
for $z\gg1$. While for $\xi>0$, the modulus field is stabilized approximately at 
\beqa
\Im\langle\tau\rangle\sim \f{11.5}{\xi}+\f{1}{3\xi}\ln(\pi\tilde{l}_N)+4.23~.
\eeqa

We note that the run away vacuum for modulus field at $i\infty$ can also be stabilized in a way similar to that of Kahler modulus stabilization with a "racetrack" type superpotential in KKLT setup~\cite{Kachru:2003aw}. In our case, because of the $\Phi^{-3}$ factor in the scalar potential, additional term in the superpotential of the form $F[H(\tau)]$ can play a similar role of $B e^{-bT}$ within the racetrack type superpotential 
\beqa 
W\supseteq A e^{-aT}+B e^{-bT}.
\eeqa
For example, for $\xi>0$, we can introduce additional superpotential term proportional to $\hat{c}_1\ll 1$ to give
\beqa
W\supseteq \Lambda^3\[ \tilde{H}(\tau)-\hat{c}_1 e^{\zeta{H}(\tau)}\]~.
\label{runaway:two}
\eeqa
It can be calculated from the scalar potential that, the scalar potential near $i\infty$ scales like
\beqa
V\sim  A[H,\tl{H}] e^{-6\xi (e^{-2\pi i\tilde{l}_N\tau})}+\hat{c}_1^2 B[H,\tl{H}] e^{(2\zeta-6\xi)(e^{-2\pi i\hat{l}_N\tau})}
-\hat{c}_1C[H,\tl{H}] e^{(\zeta-6\xi)(e^{-2\pi i\hat{l}_N\tau})}~,\nn\\
\eeqa 
with $A[H,\tl{H}],B[H,\tl{H}],C[H,\tl{H}]$ typical polynomial of $H(\tau),\tl{H}(\tau)$. For $\zeta>3\xi>0$ and $\hat{c}_1\ll 1$, it can checked that the modulus can be stabilized at some point $\langle\tau\rangle=i\tau_0$ with large real $\tau_0$.

Similarly, when $\xi=0$, we can adopt following superpotential with the additional term proportional to $\hat{c}_1\ll 1$
\beqa
W\supseteq \Lambda^3\[ \tilde{H}(\tau)-\hat{c}_1 \hat{H}(\tau)\]~.
\label{runaway:three}
\eeqa
After we calculate the scalar potential, we can see that the scalar potential scale like
\beqa
V\sim A_0 e^{2\tilde{k}_N\pi i\tau}+B_0 \hat{c}_1^2 e^{2\hat{k}_N\pi i\tau}-C_0\hat{c}_1 e^{(\tilde{k}_N+\hat{k}_N)\pi i\tau}~,
\eeqa
in the asymptotic region, with the hatted indices for $\hat{H}(\tau)$, $A_0,B_0,C_0$ some positive coefficients, and 
\beqa
\hat{k}_N&\equiv& 6\(\f{{m}}{2}+\f{{n}}{3}+{N}\)-\f{1}{2}-2\(\f{\hat{m}}{2}+\f{\hat{n}}{3}+\hat{N}\)~.
\eeqa
For $k_N\geq 0$ and $\hat{k}_N\leq 0$, the modulus can be stabilized properly at some point $\langle\tau\rangle=i\tau_0$ with large real $\tau_0$.


 For those parameter points that satisfy $H({\tau}_i)=0$, consequently $\Phi(\tau_i,\bar{\tau}_i)=0$, the scalar potential diverges at those $\tau_i$ values because of the $\[\Phi(\tau,\bar{\tau})\]^3$ factor in the denominator, which is not well defined by~(\ref{metric:constraints}). Although the scalar potential should always be stationary at those finite fixed points $\tau=i,\omega$ by symmetry arguments, such local extrema should always be local maximum instead of local minimum, except for the choices of $H(\tau)$ with which $H(i)\neq 0$ (or $H(\omega)\neq 0$).

When $\tl{H}(\tau)=0$ and at the same time $H(\tau)\neq 0$, from the extrema condition of the scalar potential in (\ref{Fderivative:V}), it is obvious that $\tl{H}^{\pr\pr}(\tau)=\tl{H}^{\pr}(\tau)=0$ are sufficient conditions for $V_{E;\tau}=0$. We have the following discussions on such conditions (see our previous work~\cite{natural:mu} for some relevant details)
\bit
\item  When $\tau=i$ is a solution of $\tl{H}(\tau)= 0$, it can be proven that $\tl{H}'(i) = 0$ can be satisfied for $\tl{m}> 1$ while $\tl{H}^{\pr\pr}(i) = 0$ can be satisfied for $\tl{m}\neq 0,2$. Similarlly, when $\tau=\omega$ is a solution of $\tl{H}(\tau)= 0$, it can be proven that $\tl{H}'(\omega) = 0$ can be satisfied for $\tl{n}> 1$ while $H^{\pr\pr}(\omega) = 0$ can be satisfied for $\tl{n}\neq 0,2$.

Therefore, given the values of $\tl{H}^{\pr\pr}(\tau)$ and $\tl{H}^{\pr}(\tau)$ at the fixed point $\tau=i,\omega$ (see Appendix \ref{appendix:B}) and at the same time non-vanishing $H(i)$ (or $H(\omega)$), it is obvious that $\tau=i$ (and $\tau=\omega$) are local extrema of the scalar potential for $\tl{m}>2$ (or $\tl{n}>2$), respectively. Such conclusions agree with the arguments by symmetry at the beginning of this section.

\item When $\tl{H}(\tilde{\tau}_i)= 0$ with $\tl{P}(j(\tilde{\tau}_i))=0$ and at the same time $H(\tilde{\tau}_i)\neq 0$, the condition $\tl{H}^{\pr\pr}(\tau)=\tl{H}^{\pr}(\tau)=0$ can be satisfied for general $\tl{P}(j(\tau))$ of the form
 \beqa
 \tl{P}(j(\tau))=\pm \tl{g}(j(\tau)) \prod\limits_{i} (j(\tau)-j(\tilde{\tau}_i))^{\tl{k}_i}~,~~~~k_i\in \mathbb{N}~
 \label{PjForm}
 \eeqa
with $\tl{k}_i \geq 3$ and $\tl{g}(j(\tau))$ an arbitrary positive polynomial function of $j(\tau)$. 
\eit

In addition to those stationary points determined by $\tl{H}(\tau)=0$ with proper $\tl{m},\tl{n}$ and $\tl{P}(j(\tau))$, other possible stationary points are determined by $V_{E,\tau}/|\tl{H}(\tau)|^2=0$ for non-vanishing $\tl{H}(\tau)$, which possibly correspond to negative vacuum energies. The positivity of the Hessian matrix as well as the sign of $V_{x_1x_1}$ at the stationary points can be determined numerically to identify various local minima, before we can identify the global minimum of the scalar potential by evaluating the potential energies for all extrema.

\section{\label{sec-4}Numerical Results}

 We select some benchmark scenarios with various choices of $\tl{H}(\tau)$ and $\Phi(\tau,\bar{\tau})$ to calculate numerically the potential energies of scalar potential (\ref{tauscalar:potential}) for all extrema and identify the global minimum. In our numerical results, blank regions in the fundamental domain correspond to those discarded nonphysical parameter points, whose potential energies exceed $M_P^4$.

 We would like to discuss several interesting scenarios:
 \bit
 \item Trivial superpotential, which amounts to $\tl{H}(\tau)=1$ in (\ref{superpotential}).
 
   In this case, the scalar potential is fully determined by the form of $\Phi(\tau,\bar{\tau})$ and $H(\tau)$. The behavior of the scalar potential is totally different from ordinary modulus stabilization studies with a non-trivial superpotential. As the scalar potential diverges at those parameter points that satisfy $H({\tau}_i)=0$, 
 finite fixed points $\tau=i,\omega$  can not be local minimums except for the case $H(i)\neq 0$ (that is, $m=0$) or $H(\omega)\neq 0$ (that is, $n=0$).

\begin{figure}[hbtp]
\begin{center}
\includegraphics[width=2.0 in]{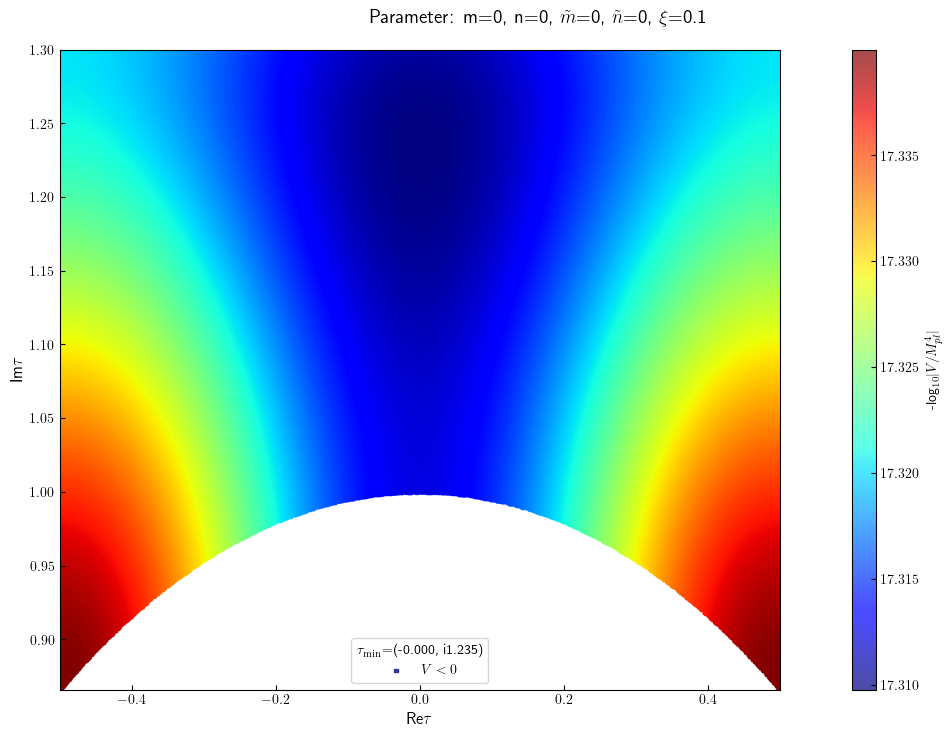}
\includegraphics[width=2.0 in]{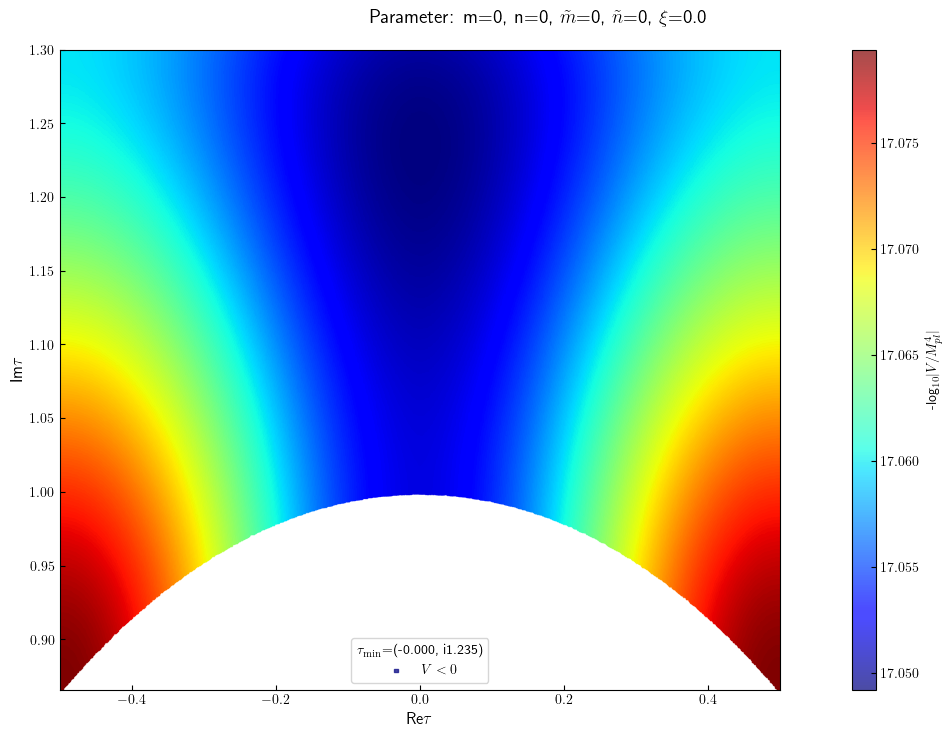}\\
\includegraphics[width=2.0 in]{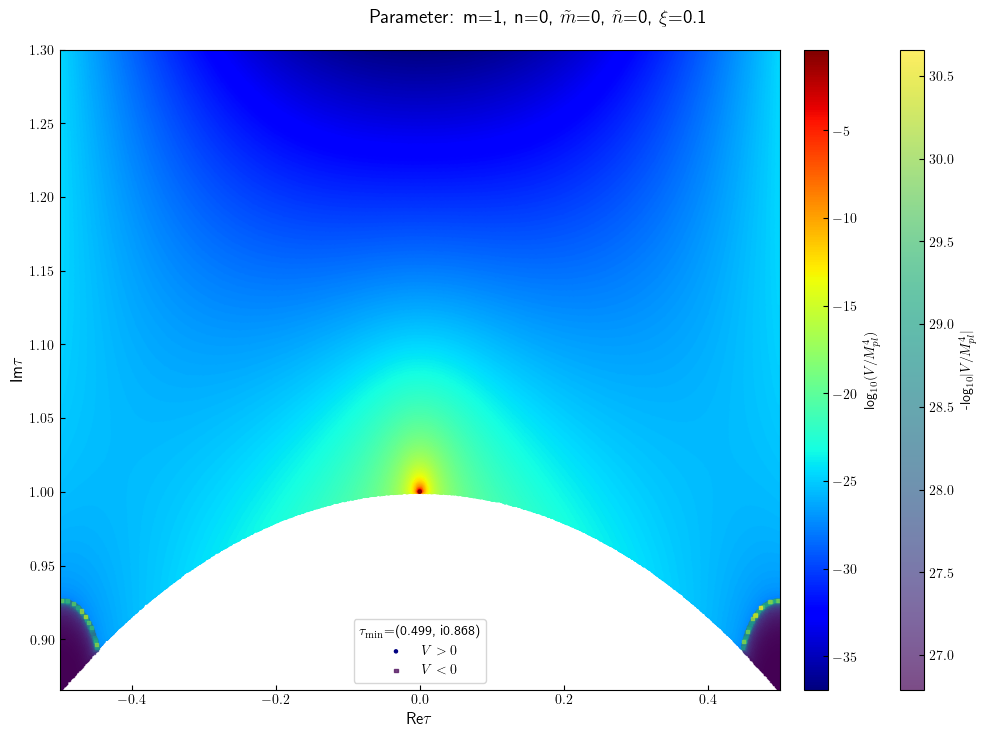}
\includegraphics[width=2.0 in]{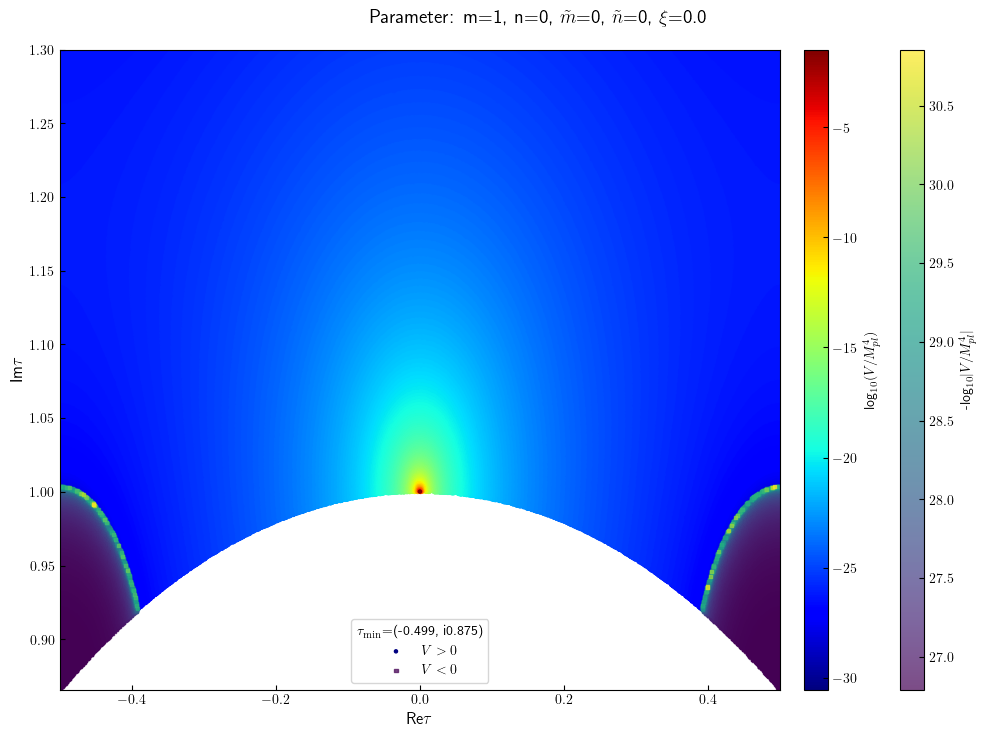}\\
\includegraphics[width=2.0 in]{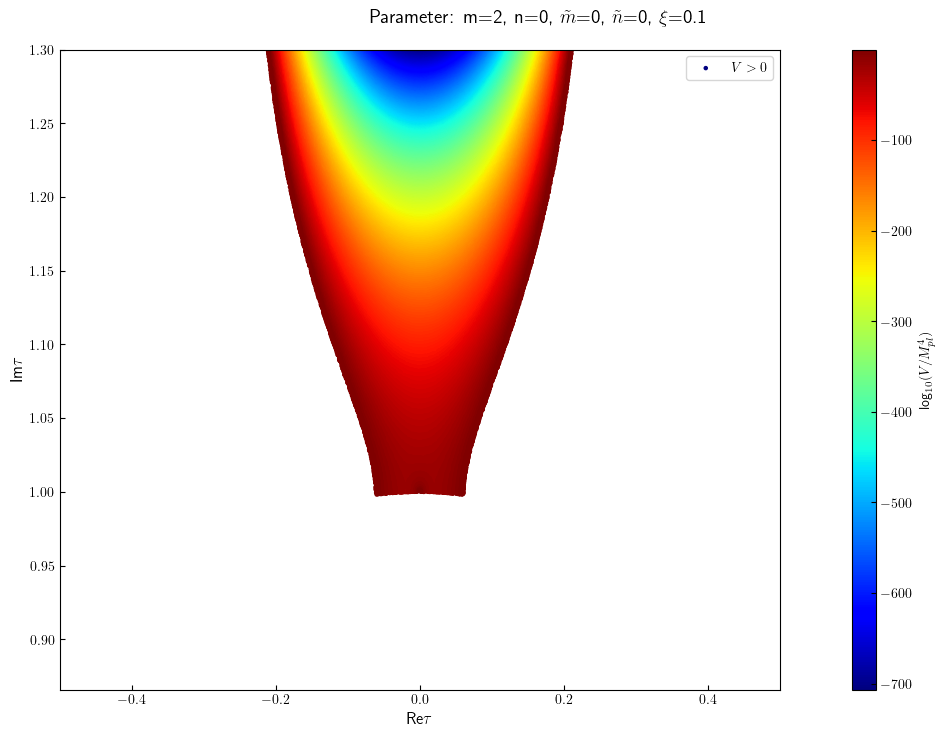}
\includegraphics[width=2.0 in]{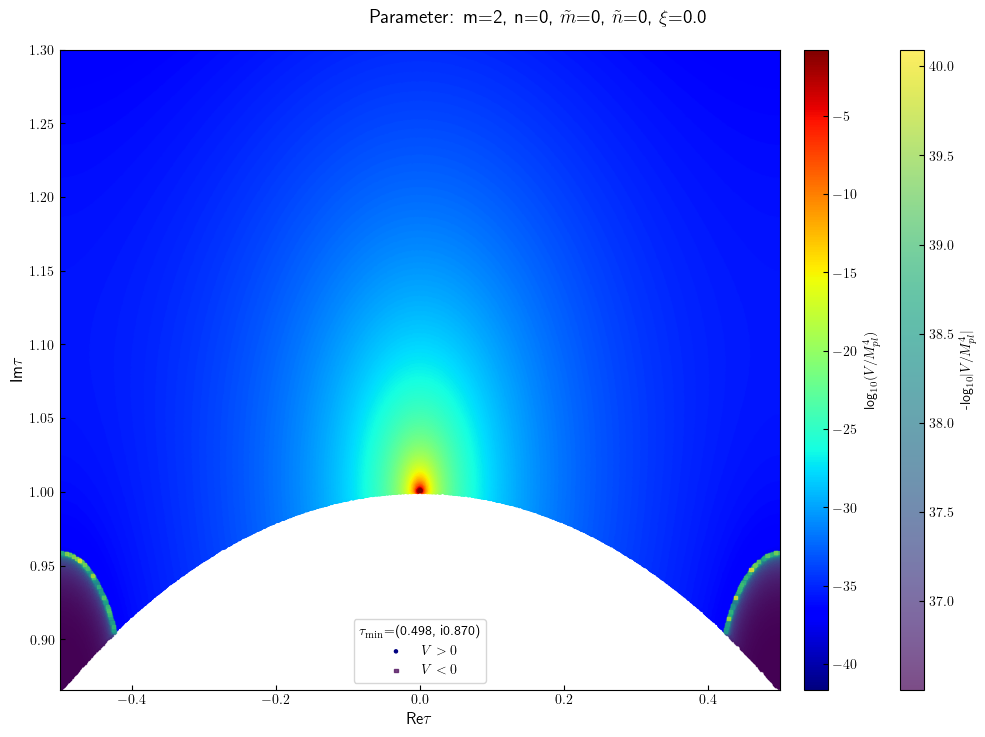}\\
\includegraphics[width=2.0 in]{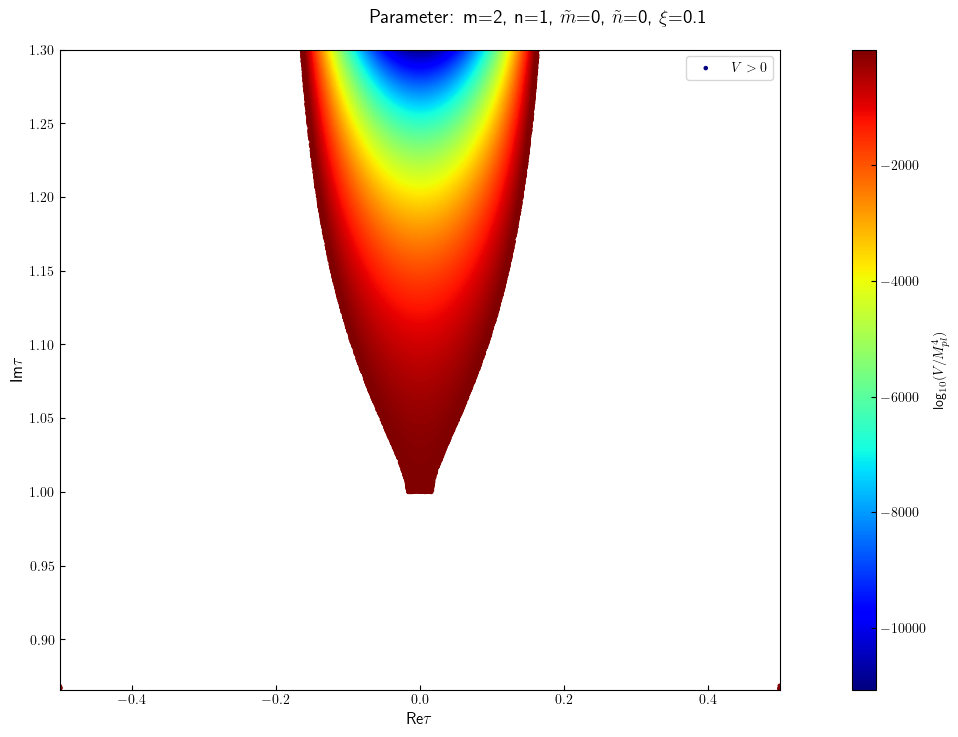}
\includegraphics[width=2.0 in]{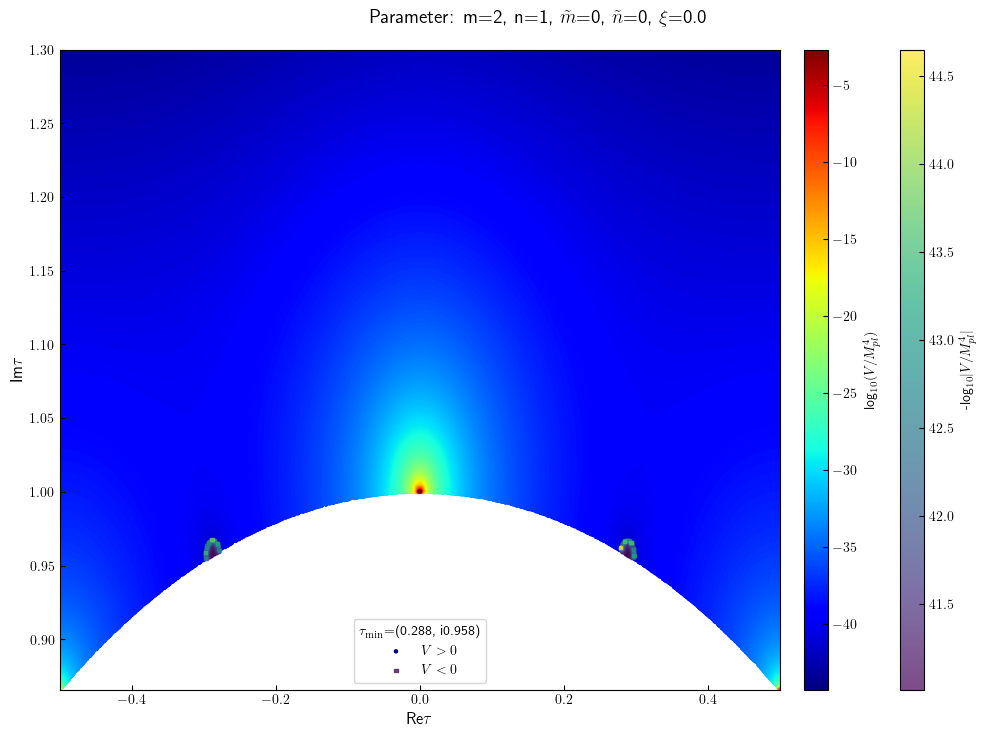}\\
\includegraphics[width=2.0 in]{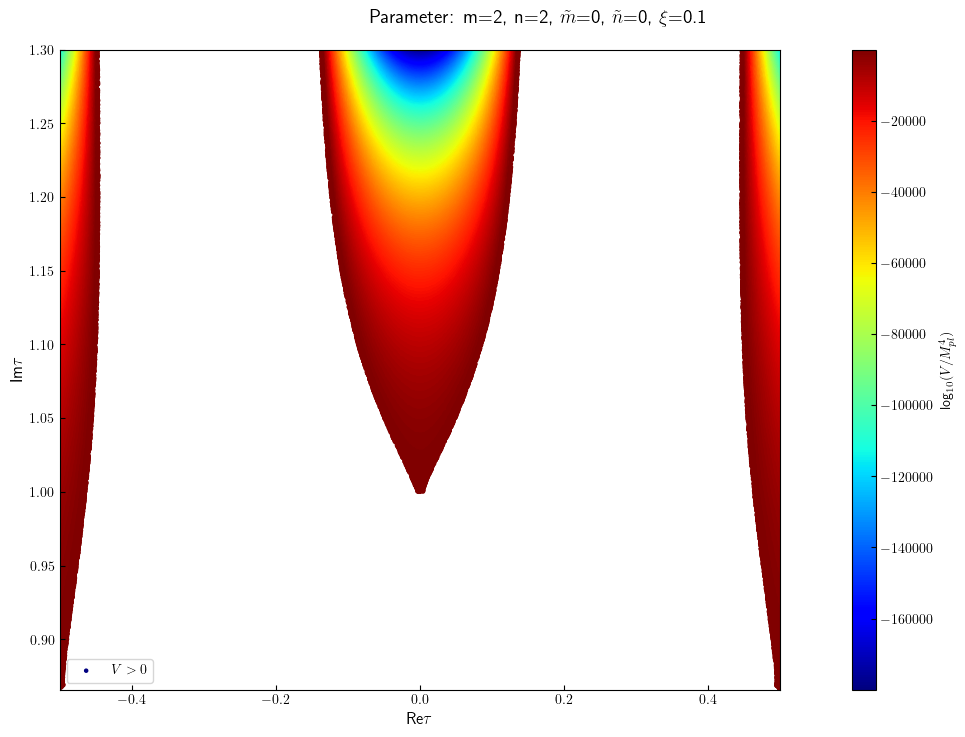}
\includegraphics[width=2.0 in]{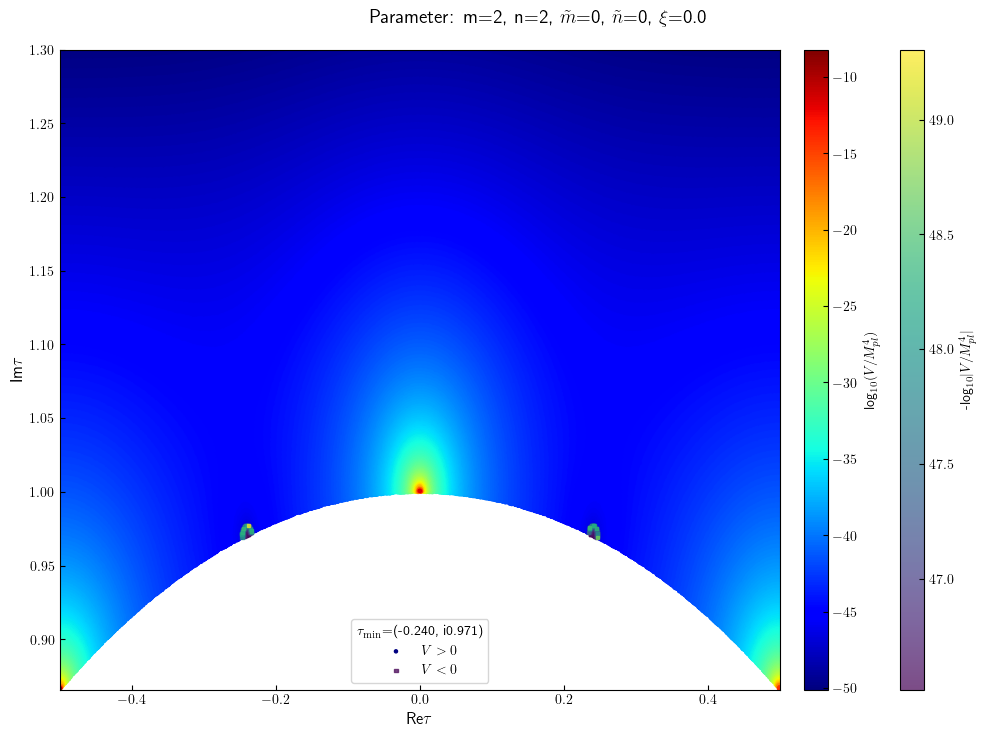}\\
\vspace{-.5cm}\end{center}
\caption{The values of $V(\tau,{\tau}^*)$ within the fundamental domain for different choices of $(m,n)$ in $H(\tau)$ with $\tl{m}=\tl{n}=0$ in the scenarios $\xi=0.1$ (left panels) and $\xi=0$ (right panels), respectively. We adopt $\mathcal P(j(\tau))=\tl{P}(j(\tau))=1$ for the all panels. We show the values of the scalar potential for $V>0$ and $V<0$ with $\log_{10}V$ and $\log_{10}|V|$ by different color bars, respectively. The location of the global minimum is also shown in each panel (or leave blank if the global minimum is the $i\infty$ fixed point). Blank regions in the fundamental domain correspond to those discarded nonphysical parameter points, whose potential energies exceed $M_P^4$.}
\label{fig1}
\end{figure}

 In Fig.\ref{fig1}, we show the potential energy for the values of the modulus field within the fundamental domain, for various choices of $({m},{n})$ with ${P}(j(\tau))=1$ and $\xi=0.1$ (or $\xi=0$). 
 From the left panels in Fig.\ref{fig1}, it can be seen that $\tau=i\infty$ is always a local extrema of the scalar potential with vanishing vacuum energy for $\xi>0$ (although the fundamental domain is not fully displayed in the panels). When $\xi=0$, the infinite fixed point $\tau=i\infty$ is also a local extrema of the scalar potential for $m+n\neq 0$, as the condition (\ref{minimum:iinfty}) is always fulfilled for non-trivial $H(\tau)$. 
 

It can be seen from the upper left panel that the global minimum of the scalar potential lies at about $\tau=1.235i$ on the imaginary axis, which slightly deviate from the minimum $\tau\simeq 1.2i$ obtained with the same form of $H(\tau)$ that studied in~\cite{NPP:2201.02020} (with non-trivial superpotential). Turning off the $\xi$ parameter (however, keeping the $|H(\tau)|^2$ term in the frame function) will still lead to a tiny deviation from the previous minimum $\tau\simeq 1.235i$. Similarly, when $\xi=0$, the global minimum of the scalar potential (see the panels correspond to $(m,n)=(1,0)$, $(2,1)$ and $(2,2)$) can be checked to deviate slightly from the minimum obtained with the same form of $H(\tau)$ that given in~\cite{NPP:2201.02020}.  When $m+n\geq 2$ for $\xi>0$, the infinite fixed point $\tau=i\infty$ is always the global minimum of the scalar potential, leading to a runaway type vacuum. Such a minimum is totally different from the minimum obtained with the same form of $H(\tau)$ given in~\cite{NPP:2201.02020}.
It should be noted that the global minimum for $(m,0)$ case still lead to CP-breaking vacuum, which slightly depart from the left (right) cusp symmetric point and the boundary (for example, see the $(m,n)$=$(1,0)$,$(2,0)$ panels). Such minimums can be phenomenologically interesting, as the origin of CP violation can be explained in addition to the flavor structure.

 \item  $\tl{H}(\tau)$ takes the same form as $H(\tau)$.
 
\begin{figure}[hbtp]
\begin{center}
\includegraphics[width=2.0 in]{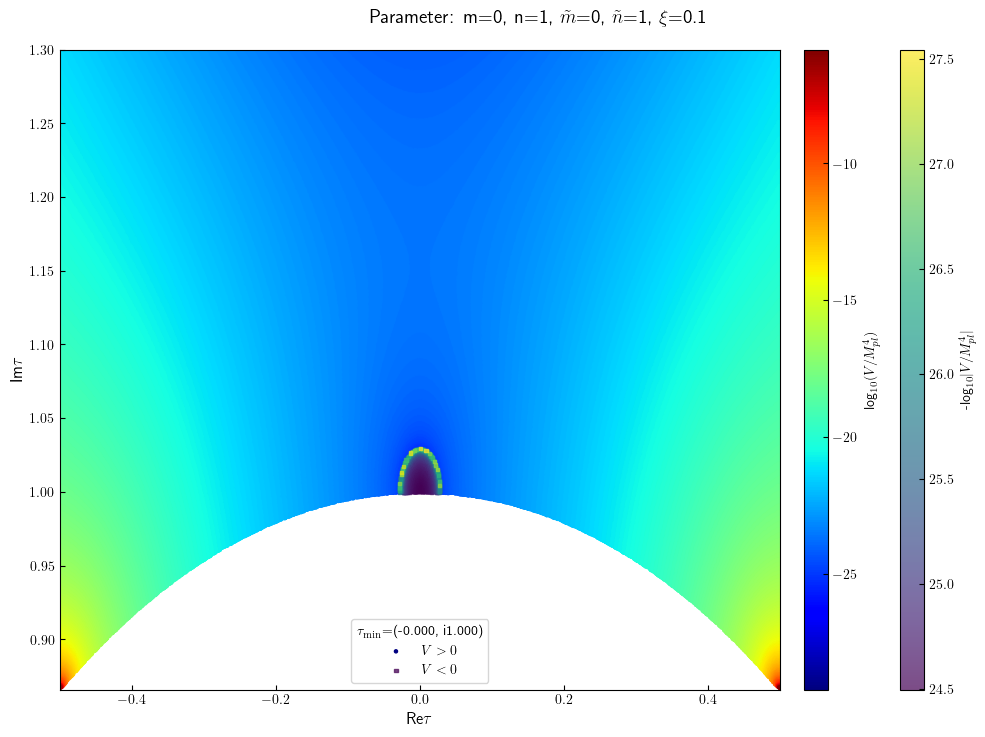}
\includegraphics[width=2.0 in]{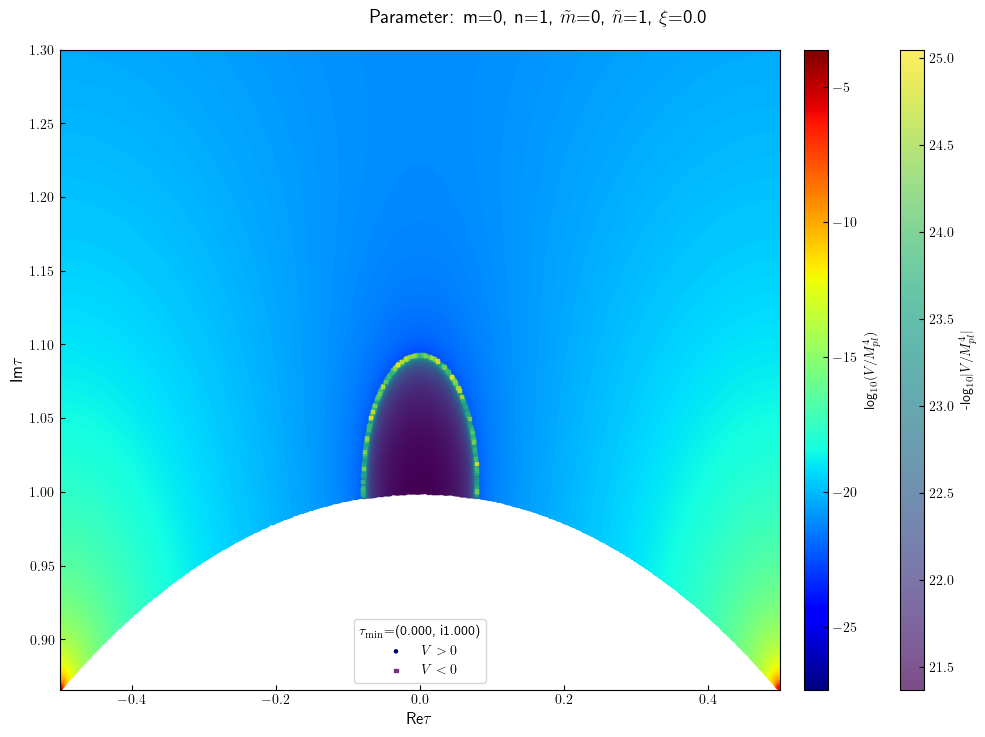}\\
\includegraphics[width=2.0 in]{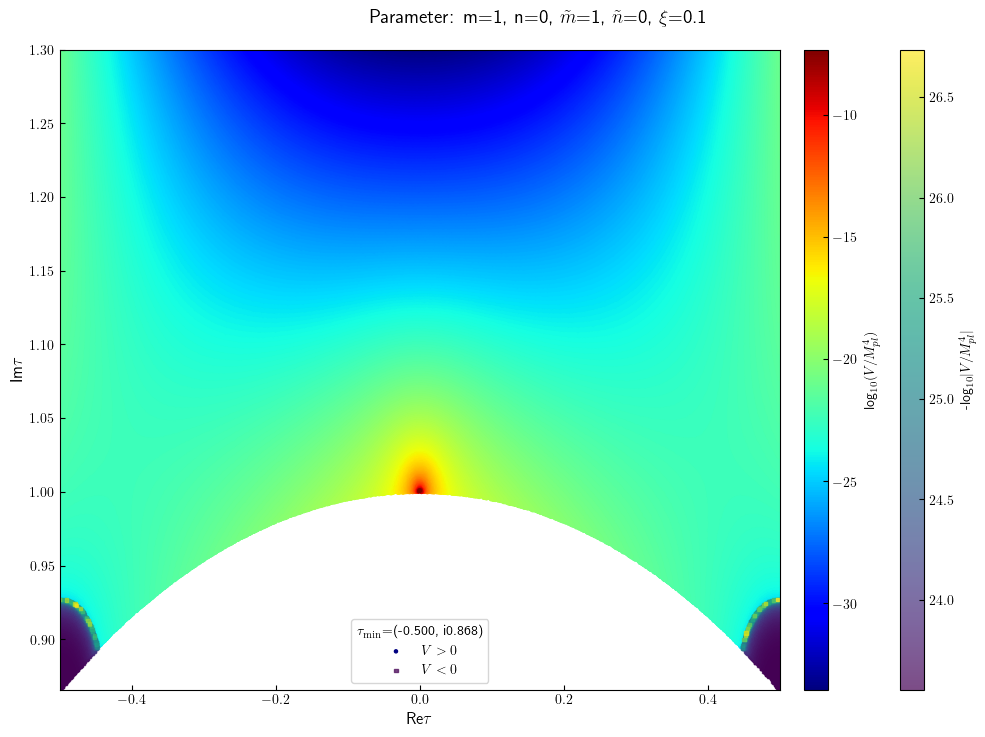}
\includegraphics[width=2.0 in]{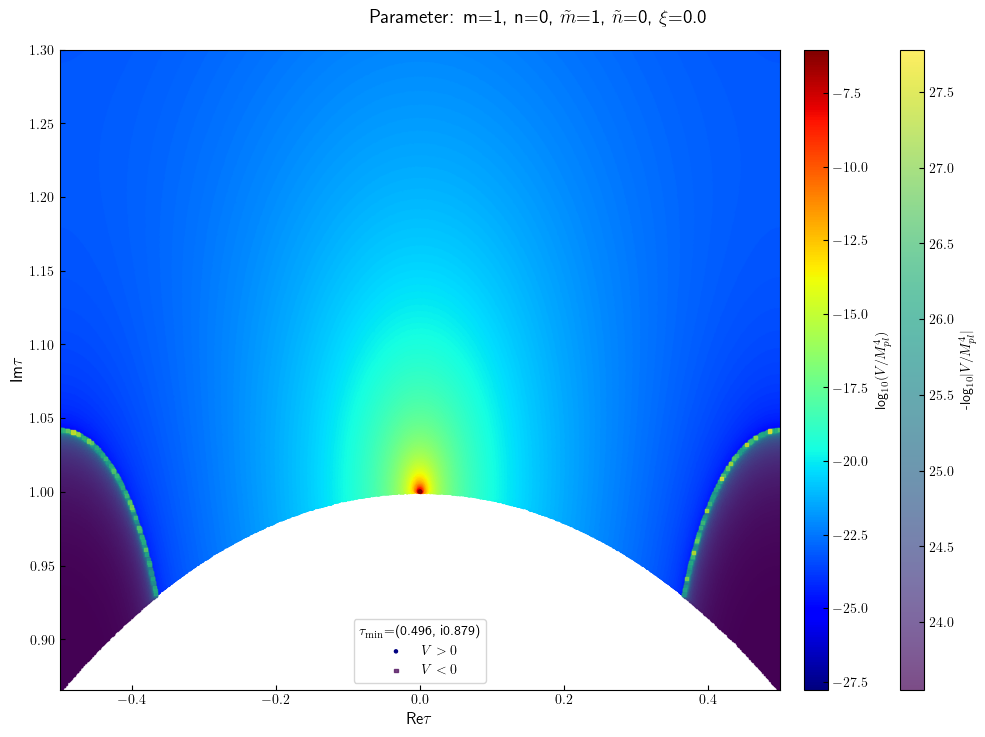}\\
\includegraphics[width=2.0 in]{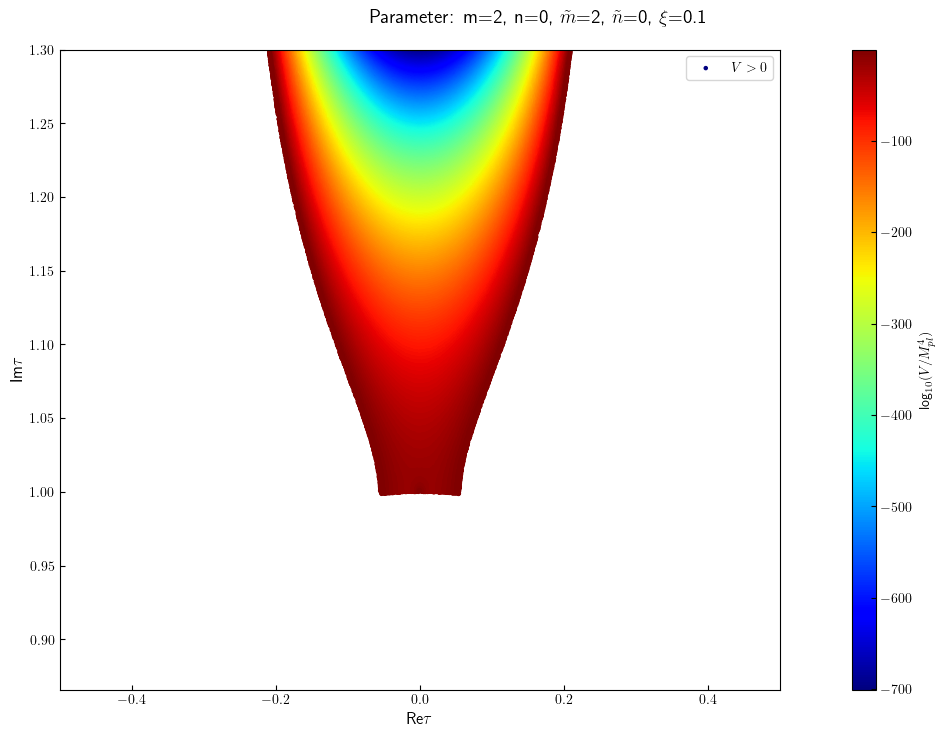}
\includegraphics[width=2.0 in]{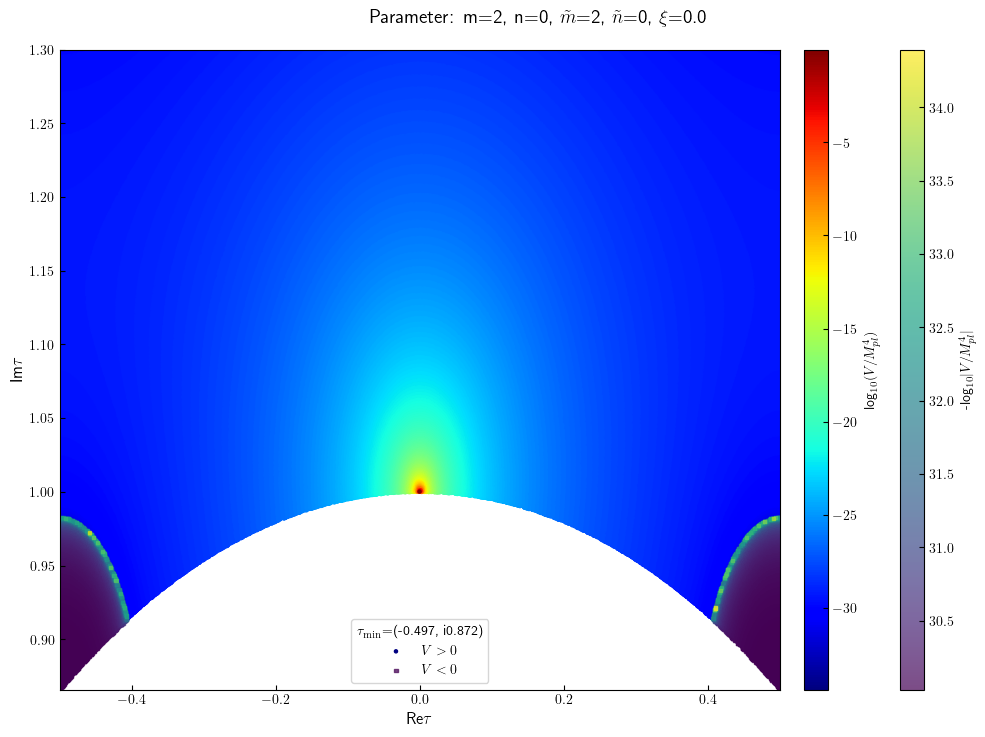}\\
\includegraphics[width=2.0 in]{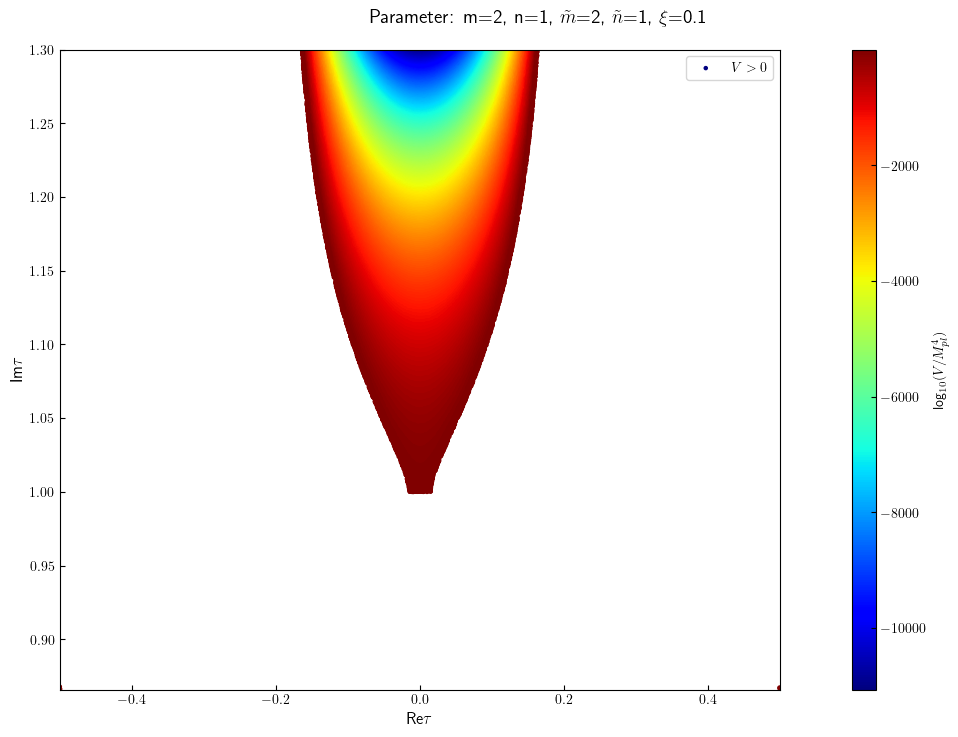}
\includegraphics[width=2.0 in]{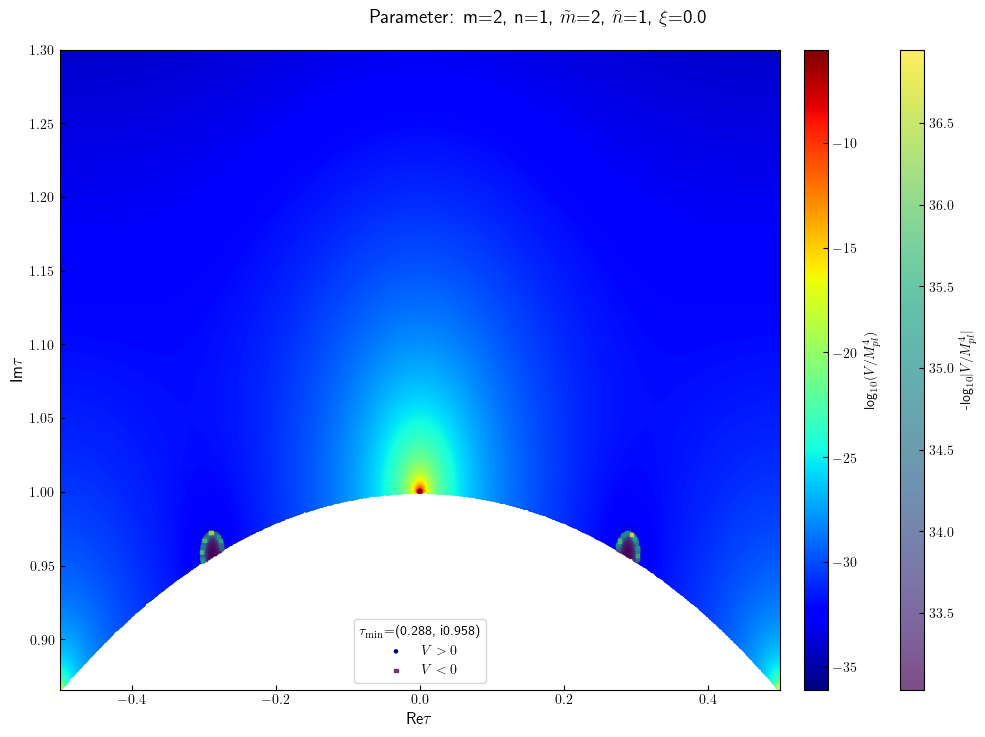}\\
\includegraphics[width=2.0 in]{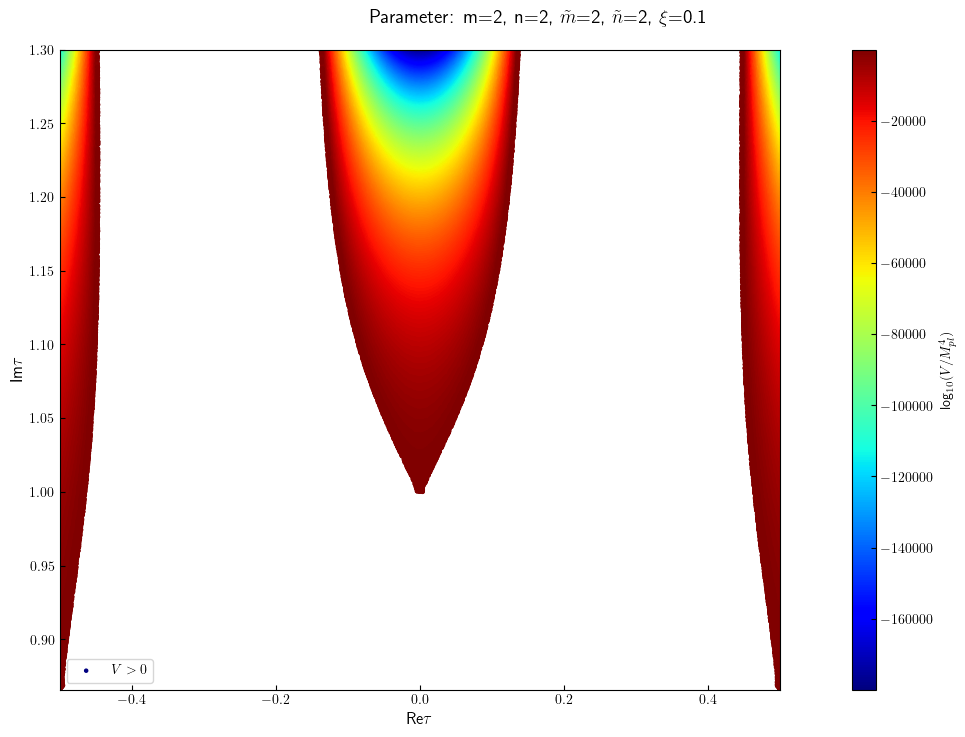}
\includegraphics[width=2.0 in]{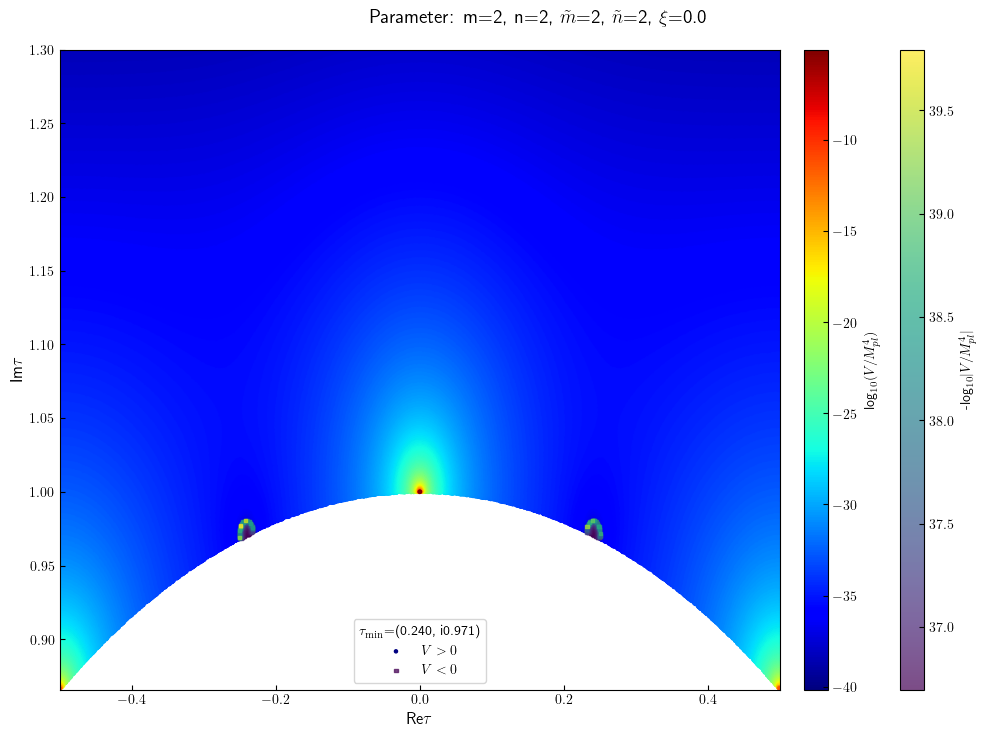}\\
\vspace{-.5cm}\end{center}
\caption{The values of $V(\tau,{\tau}^*)$ within the fundamental domain for the case $\tl{H}(\tau)$ taking the same form as $H(\tau)$ in the scenarios $\xi=0.1$ (left panels) and $\xi=0$ (right panels), respectively. Other settings are the same as that of the previous figure. }
\label{fig2}
\end{figure}

 In this case, the scalar potential is still determined by the form of $\Phi(\tau,\bar{\tau})$ and $H(\tau)$, which simplify greatly the discussions of (\ref{Fderivative:V}). 
 In Fig.\ref{fig2}, we show the potential energy for the values of the modulus field within the fundamental domain, for various choices of $({m},{n})$ with ${P}(j(\tau))=1$ and $\xi=0.1$ (or $\xi=0$). It can be proven that the scalar potential diverges at the finite fixed points $\tau=i$ (or $\tau=\omega$) except for $m=0$ (or $n=0$).

 From both Fig.\ref{fig1} and Fig.\ref{fig2}, it is clear that the panels with the same $(m,n)$ and $\xi$ choices show similar patterns. Therefore, the $H(\tau)$ appeared in the frame function $\Phi(\tau,\bar{\tau})$ determine the behavior of the scalar potential.

 Again, from the condition (\ref{minimum:iinfty}), $\tau=i\infty$ is always a local minimum of the scalar potential with vanishing vacuum energy for $\xi\geq 0$ (except for trivial $H(\tau)$ when $\xi=0$). When $m+n\geq 2$ for $\xi>0$, the infinite fixed point $\tau=i\infty$ is always the global minimum of the scalar potential, leading to a runaway type vacuum.

 Similar to the discussions in the previous scenario, the location of the global minimum is shifted slightly from the minimum obtained with the same form of $H(\tau)$ that studied in~\cite{NPP:2201.02020}.
 Besides, the CP-breaking global minimum for $(m,0)$ case still exists, which slightly depart from the left (right) cusp symmetric point and the boundary.

\item  General choice of $\tl{H}(\tau)$ and $H(\tau)$, which adopts different choice of $(\tl{m},\tl{n})$ and $(m,n)$ (assuming $P(j(\tau))=\tl{P}(j(\tau))=1$).
 
\begin{figure}[hbtp]
\begin{center}
\includegraphics[width=2.0 in]{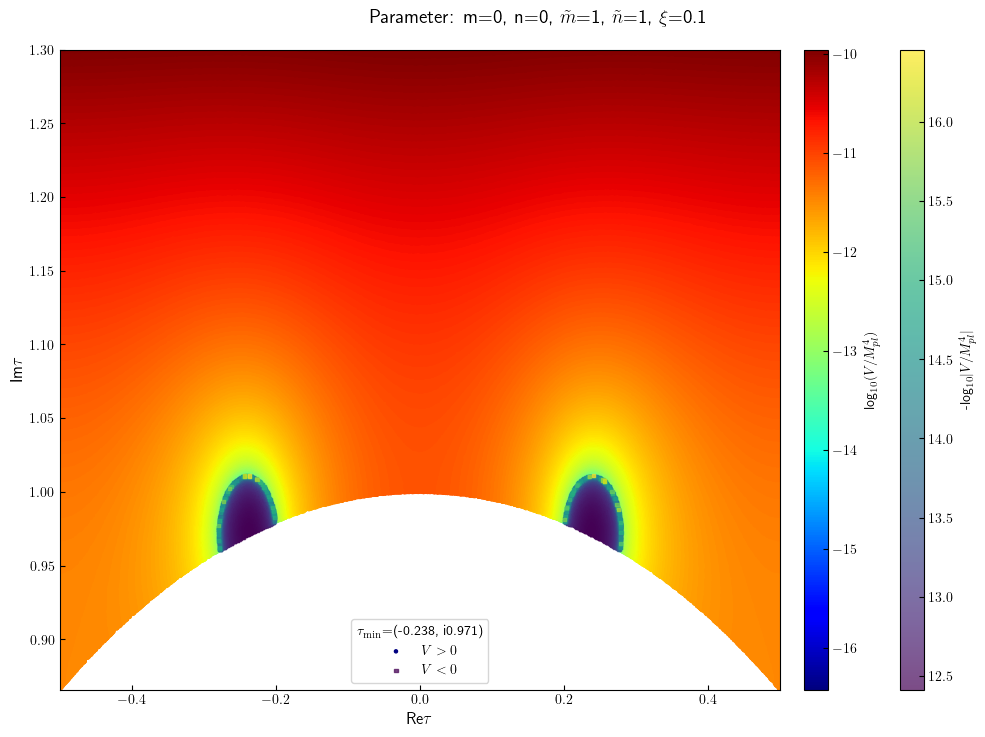}
\includegraphics[width=2.0 in]{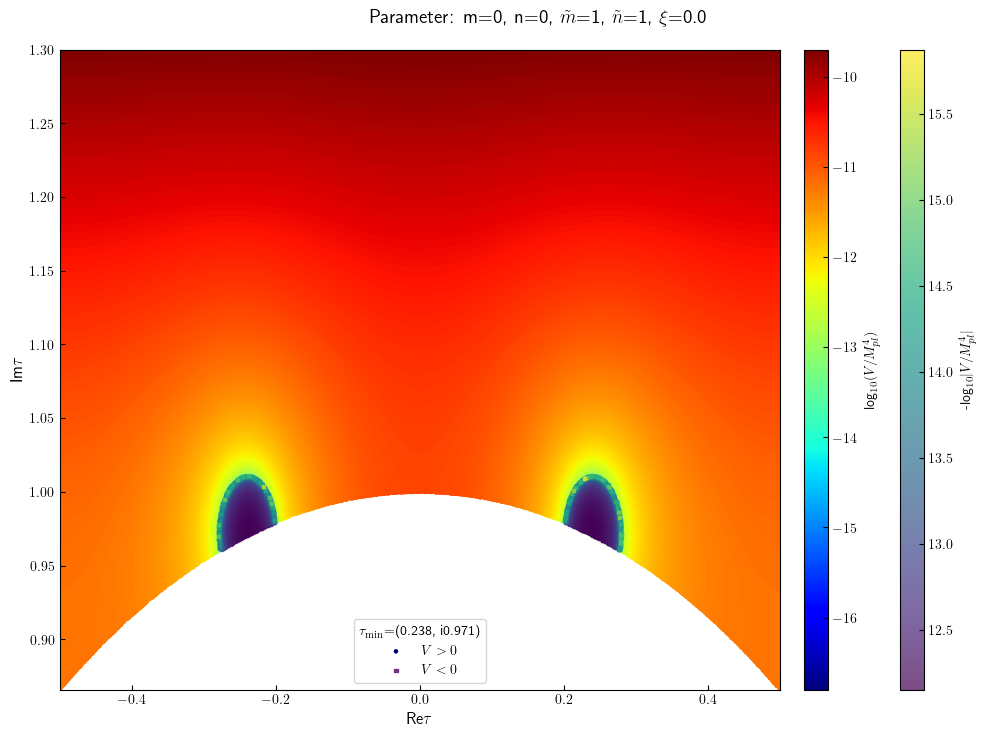}\\
\includegraphics[width=2.0 in]{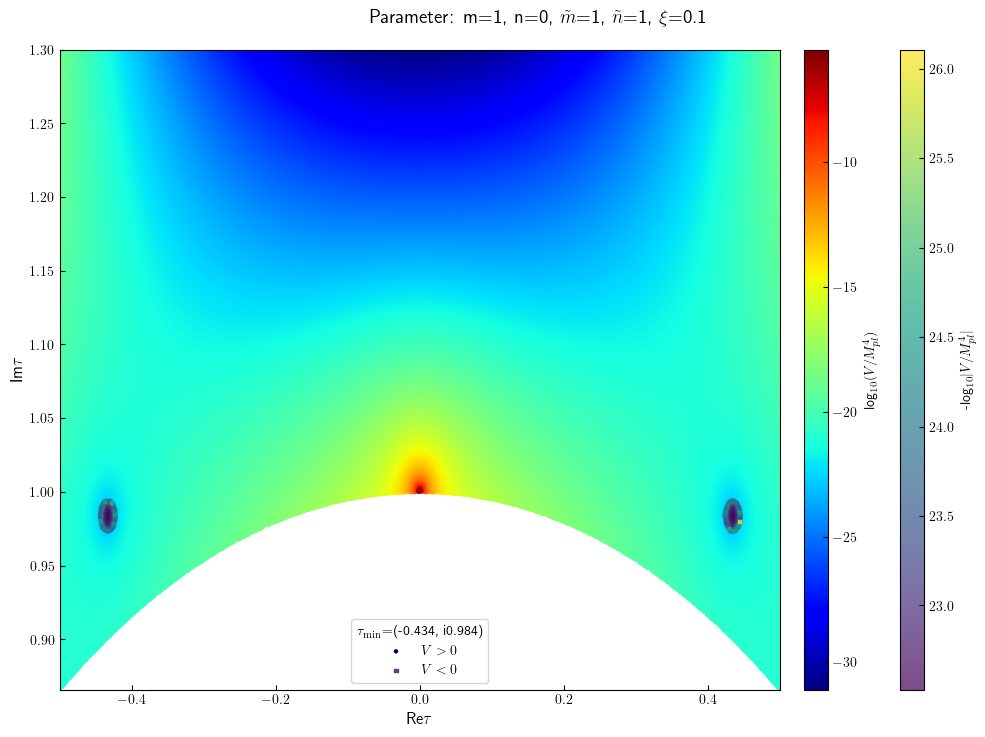}
\includegraphics[width=2.0 in]{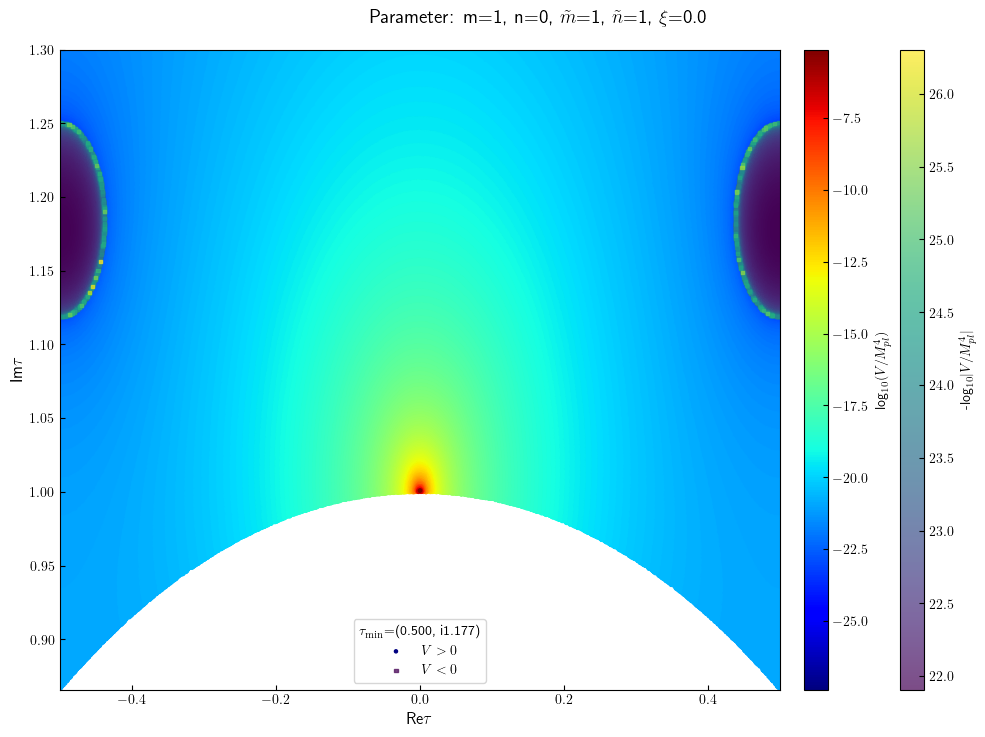}\\
\includegraphics[width=2.0 in]{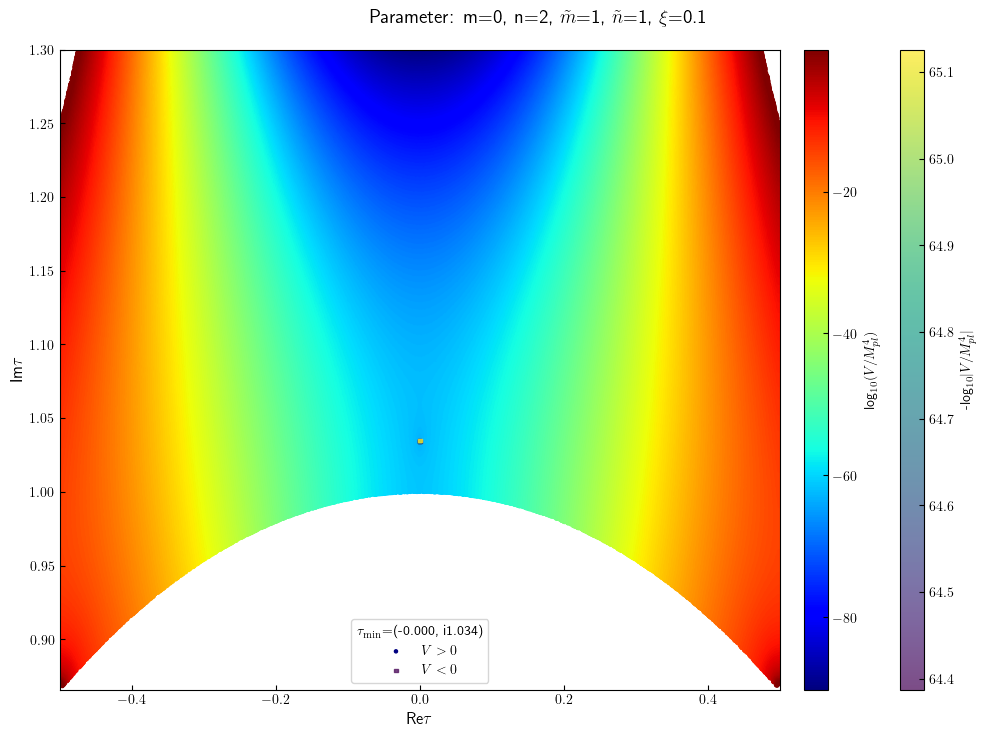}
\includegraphics[width=2.0 in]{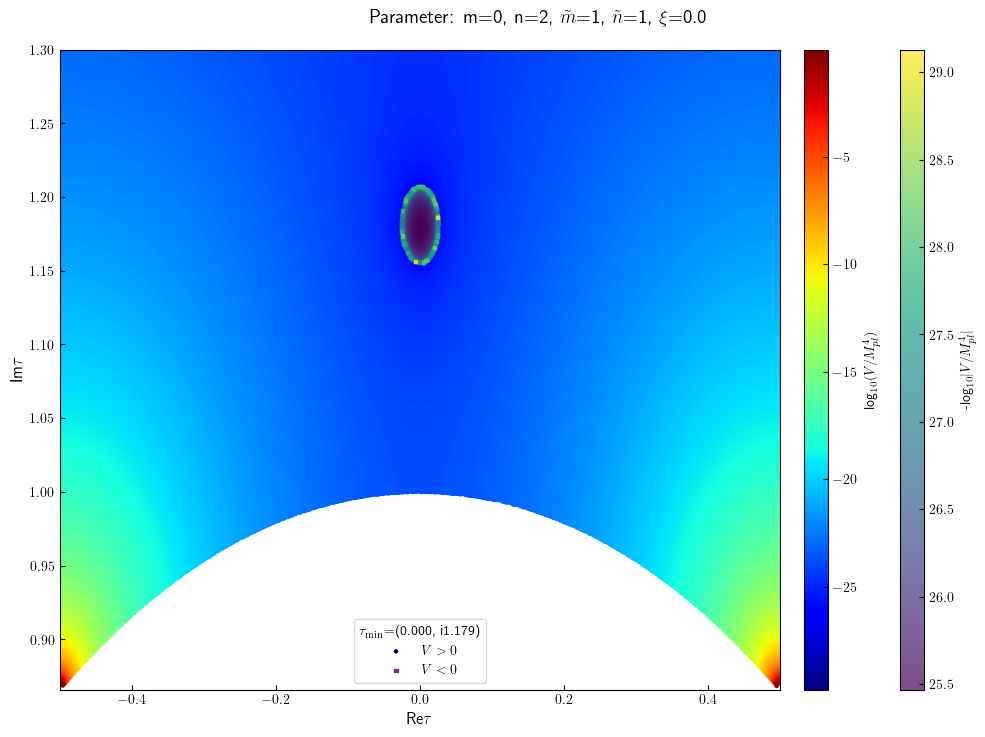}\\
\includegraphics[width=2.0 in]{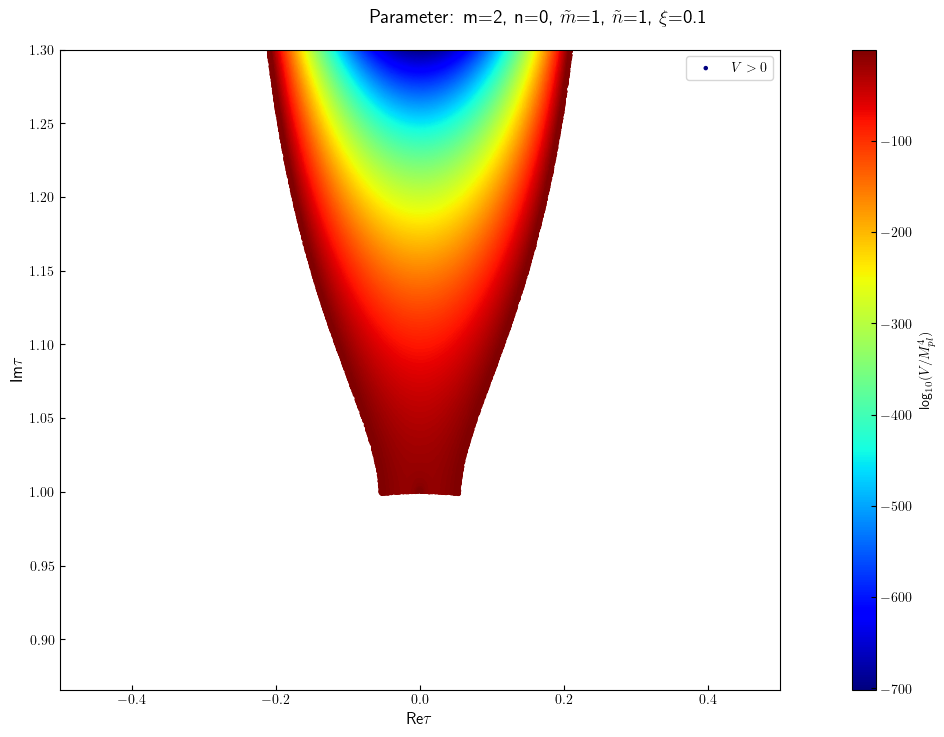}
\includegraphics[width=2.0 in]{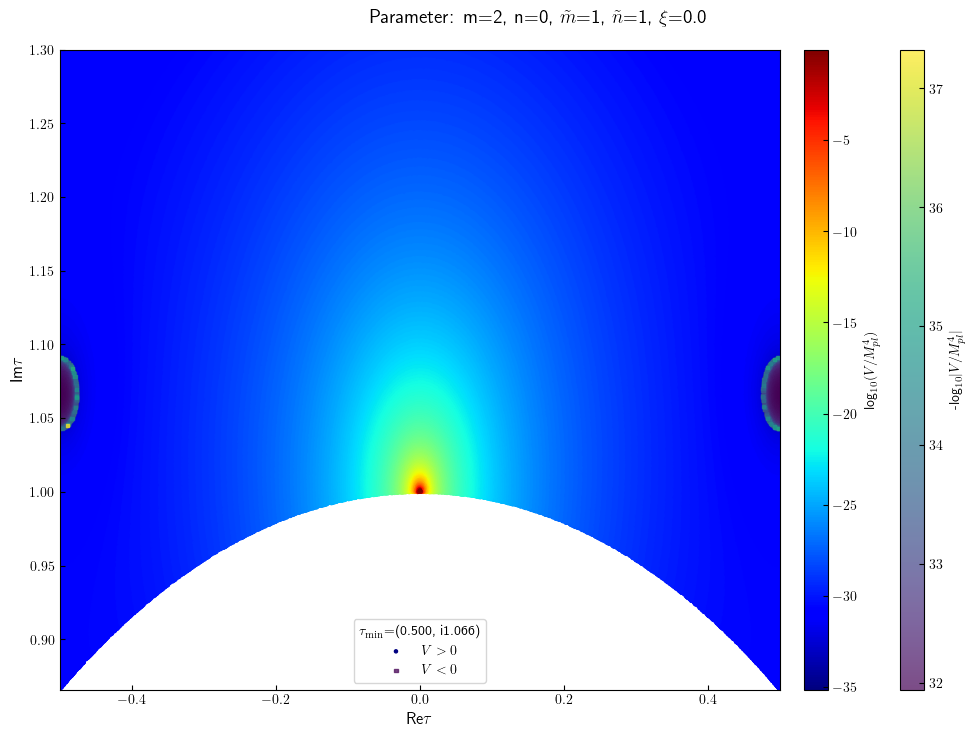}\\
\includegraphics[width=2.0 in]{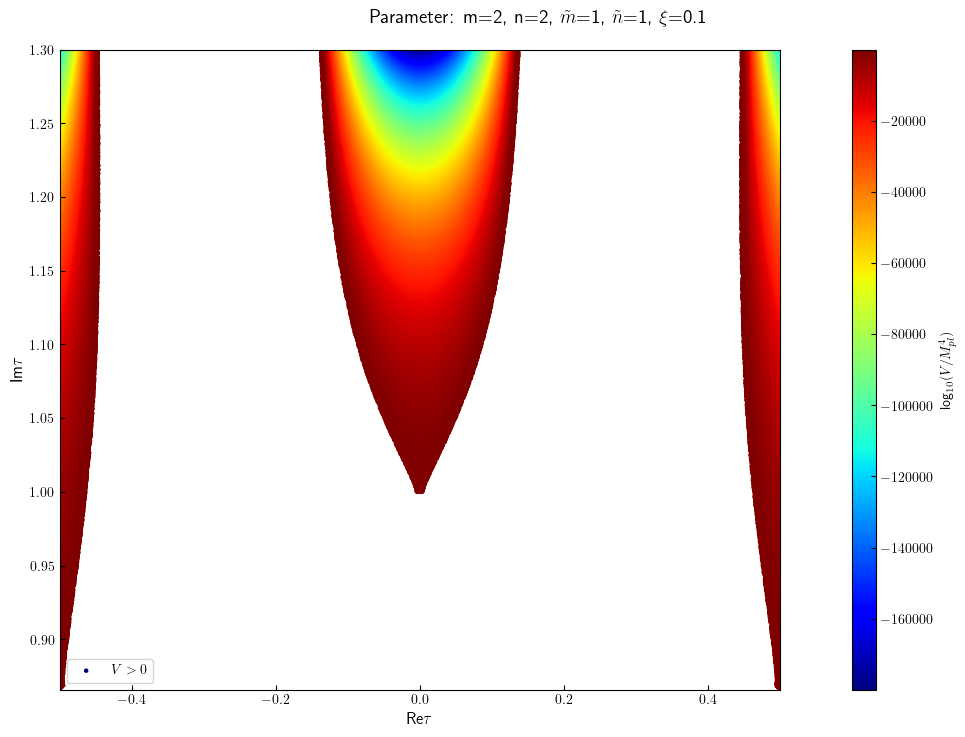}
\includegraphics[width=2.0 in]{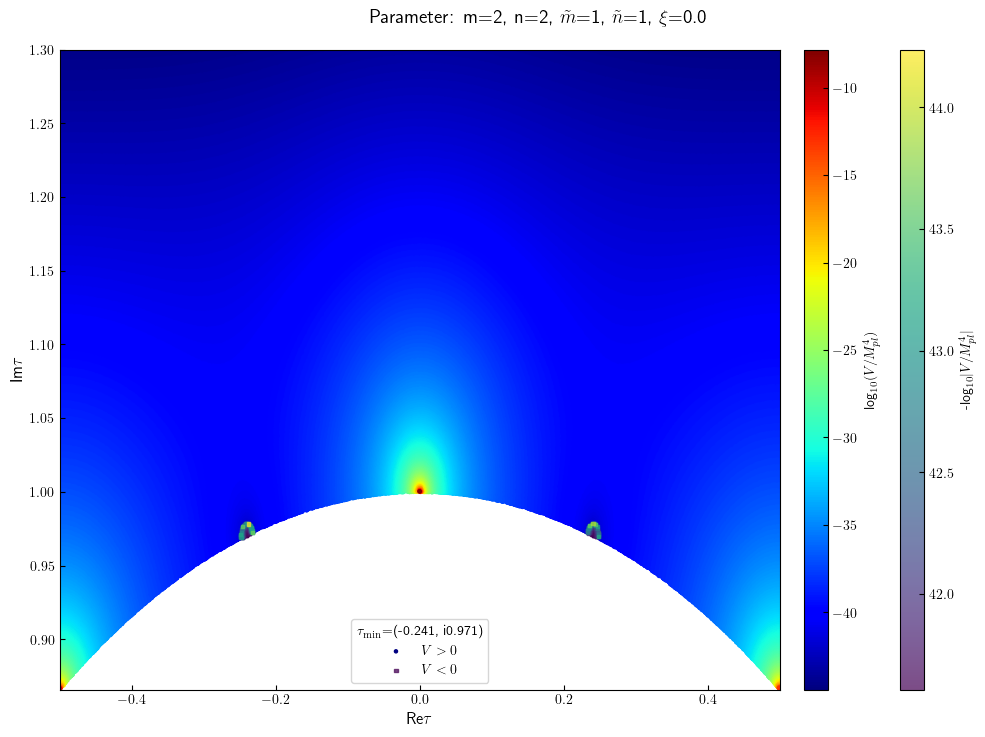}\\
\vspace{-.5cm}\end{center}
\caption{The values of $V(\tau,{\tau}^*)$ within the fundamental domain for the case $\tl{H}(\tau)$ taking different form from $H(\tau)$ in the scenarios $\xi=0.1$ (left panels) and $\xi=0$ (right panels), respectively. Other settings are the same as that of the previous figures. }
\label{fig3}
\end{figure}

 In this case, the scalar potential is determined by both the form of $\Phi(\tau,\bar{\tau})$ and $\tl{H}(\tau)$.
 In Fig.\ref{fig3}, we show the potential energy for the values of the modulus field within the fundamental domain, for various choices of $({m},{n}), (\tl{m},\tl{n})$ with ${P}(j(\tau))=\tl{P}(j(\tau))=1$ and $\xi=0.1$ (or $\xi=0$). Unlike the previous cases, the panels with the same $\xi$ choices and the same $(m,n)$ (or $(\tl{m},\tl{n})$) show different patterns, which is interesting phenomenologically. The scalar potential still diverges at the finite fixed points $\tau=i$ (or $\tau=\omega$) except for $m=0$ (or $n=0$).

  From the panels in Fig.\ref{fig3}, it can be seen that $\tau=i\infty$ is always a local extrema of the scalar potential with vanishing vacuum energy for $\xi>0$ when $H(\tau)$ is non-trivial. When $\xi=0$, the infinite fixed point $\tau=i\infty$ is also a local extrema of the scalar potential when the condition (\ref{minimum:iinfty}) is fulfilled.

 From the panels, it can be seen that new location of CP-breaking global minimum, for example,  $\tau=-0.434+0.984i$ for $(m,n)=(1,0)$ and $ (\tl{m},\tl{n})=(1,1)$ when $\xi=0.1$, can be realized.

 From previous discussions, we know that finite fixed point $\tau=i$ (or $\omega$) is a local extrema of the scalar potential for $\tl{m}>2$ (or $\tl{n}>2$) when $H(i)\neq 0$ (or $H(\omega)\neq 0$) and $\tilde{H}(i)=0$ (or $\tilde{H}(\omega)=0$) . 
 The corresponding vacuum energies at such finite fixed points $\tau=i$ (or $\omega$) always vanish because $\tilde{H}^\pr(i)=\tilde{H}^{\pr\pr}(i)=0$ (or $\tilde{H}^\pr(\omega)=\tilde{H}^{\pr\pr}(\omega)=0$), leading to a Minkowski local (or global) minimum. 
  If the $\tau=i\infty$ runaway vacuum can be uplifted by the stabilization mechanism discussed previously in (\ref{runaway:one}), (\ref{runaway:two}) and (\ref{runaway:three}), the modulus can be stabilized at some large value at the imaginary axis ($\Im\tau\gg 1$) near $\tau=i\infty$. Modulus VEV with large value at the imaginary axis can be fairly interesting phenomenologically. Such a vacuum with tiny positive vacuum energy can be very long-lived if the global minimum located elsewhere is of Minkowski type. This meta-stable de-Sitter type vacuum can possibly be compatible with current cosmological observations.

\item  Most general choices of $\tl{H}(\tau)$ and $H(\tau)$ with non-trivial $P(j(\tau))$ and $\tl{P}(j(\tau))$.

We show in Table.\ref{most:general} the global minimum of the scalar potential for some general choices of $\tl{H}(\tau)$ and $H(\tau)$ with non-trivial $P(j(\tau))$ and $\tl{P}(j(\tau))$. 
It can be seen from the Table that new locations of CP-breaking global minimum can easily be realized by adopting non-trivial choices of $P(j(\tau))$ and $\tl{P}(j(\tau))$.

\begin{table}[htbp]
\centering
\begin{tabular}{|c|c|c|c|c|c|c|c|c|}
\hline
$m$ & $n$ & $\tilde{m}$ & $\tilde{n}$ & $\xi$ & $P(j(\tau))$ & $\tilde{P}(j(\tau))$ & $\tau_{\mathrm{min}}$ & $V_{\mathrm{min}}/M_{\mathrm{pl}}^{4}$ \\
\hline
0 & 1 & 0 & 1 & 0.1 & $ [j(\tau)-j(2i)]^{2} $ & 1 & $0.448 + 1.223i$& $ 0.0 $ \\ 
\hline
1 & 0 & 0 & 0 & 0 & 1 & $ [j(\tau)-j(2i)]^{2} $ & $-0.497+0.874i$& $-1.103\times 10^{-5} $ \\ 
\hline
1 & 2 & 1 & 2 & 0.1 & $ [j(\tau)-j(0.1+2i)]^{3} $ & $ [j(\tau)-j(0.1+2i)]^{3} $ & $0.1743 +1.143i$& 0.0 \\
\hline
\end{tabular}
\caption{The minimums of the scalar potential for some benchmark points with the most general choices of $\tl{H}(\tau)$ and $H(\tau)$, adopting non-trivial $P(j(\tau))$ and $\tl{P}(j(\tau))$.}
\label{most:general}
\end{table}

 \eit 
\subsection{Hierarchical flavor structure with large $\Im\tau$} 
The hierarchies among the charged leptons and quarks masses can follow solely from the properties of the modular forms (strictly speaking, the representation of the mass term bilinear with respect to the residual group) in the vicinity of fixed points~\cite{Novichkov:2021evw,Feruglio:2021dte}, avoiding the fine-tuning of the constant parameters (or avoiding the introduction of extra weighton-like scalar fields). For a modular invariant bilinear of the form $\psi_i^c M_{ij}(\tau)\psi_j$, where $M_{ij}(\tau)$ is a modular form of level $N^\pr$ and weight $k$, the zero entries of $M_{ij}(\tau)$ at a fixed point in the mass matrices will generically become non-zero when $\tau$ deviates slightly from the fixed point. The magnitudes of these residual symmetry breaking entries are controlled by the departure $\epsilon$ of modulus VEV from that fixed point and the field transformation properties under the residual symmetry group, which may depend on the modular weights~\cite{Petcov:2023vws,Novichkov:2021evw,Petcov:2022fjf}. That is, the entries of the mass matrices in the vicinity of the fixed point are subjected to corrections of ${\cal O}(\epsilon^l)$, where the powers $l$ corresponding to the representations of the residual symmetry group and depends only on how the representations of the fermion fields in the mass term bilinear decompose under the considered residual symmetry group. The degree of suppression, given by the integer $l$, can take values
$l = 0,1,\cdots,N^\pr-1$ in the case of $\tau_{sym} = i\infty$. Residual symmetry representations for different choices of $\Ga^\pr_{N^\pr}$ (for
levels $N^\pr \leq 5$) representations are summarized in~\cite{Novichkov:2021evw}.

The hierarchies for the charged lepton masses, Pontecorvo-Maki-Nakagawa-Sakata (PMNS) mixing and neutrino masses can be explained by typical value of $\tau$ near the $i\infty$ fixed point in the modular $A_5^\pr, S_4^\pr$ model~\cite{Novichkov:2021evw}. For example, one can assign the representation with respect to $\Ga^\pr_{N^\pr}$ and modular weight quantum number $({\bf r}, k_{\bf r})$ for the three generations of lepton doublets $L_{a}\sim ({\bf 3}, 3)$, the charged-lepton singlets $E_{a}^c\sim ({\bf 3^\pr},1)$, two gauge singlets neutrinos $N^c\sim ({\bf \hat{2}^\pr},2)$ and $H_{u,d}\sim ({\bf 1},0)$ in the modular $A_5^\pr$ model. The relevant part in the frame function is given by
\beqa
\Phi_{total}\supset \f{L_{\bf 3}^\da L_{\bf 3}}{(2\Im\tau)^{3}}+\f{(E_{\bf 3^\pr}^c)^\da E_{\bf 3^\pr}^c}{(2\Im\tau)}+\f{(N_{\bf \hat{2}^\pr}^c)^\da N_{\bf \hat{2}^\pr}^c}{(2\Im\tau)^2}+ H_u^\da H_u+H_d^\da H_d~,
\eeqa
and the superpotential is given as
\beqa
    W &=& \left[
      \alpha_1 \left( Y^{(5,4)}_{\mathbf{4}} E^c L \right)_{\mathbf{1}} +
      \alpha_2 \left( Y^{(5,4)}_{\mathbf{5}, 1} E^c L \right)_{\mathbf{1}} +
      \alpha_3 \left( Y^{(5,4)}_{\mathbf{5}, 2} E^c L \right)_{\mathbf{1}}
    \right] H_d \nn \\
    &+& \left[
      g_1 \left( Y^{(5,5)}_{\mathbf{\hat{6}}, 1} N^c L \right)_{\mathbf{1}} +
      g_2 \left( Y^{(5,5)}_{\mathbf{\hat{6}}, 2} N^c L \right)_{\mathbf{1}} +
      g_3 \left( Y^{(5,5)}_{\mathbf{\hat{6}}, 3} N^c L \right)_{\mathbf{1}}
    \right] H_u \nn\\
    &+& \Lambda \left( Y^{(5,4)}_{\mathbf{3}'} (N^c)^2 \right)_{\mathbf{1}} \,.
\eeqa

 The charged-lepton mass hierarchy $(m_\tau:m_\mu:m_e) \sim (1:\epsilon:\epsilon^4)$ and PMNS mixing can be explained with $\epsilon=e^{-2\pi \Im\tau/5}$ for $\tau\simeq -0.47+3.11i$~\cite{Novichkov:2021evw}. From eq.(\ref{xizerotau:VEV}), such a stabilized $\tau$ value can be obtained by choosing $\tl{k}_N=19/6$ for the parameters $'m,n,\tl{m},\tl{n}'$ in $H(\tau)$ and $\tl{H}(\tau)$.

For the quark sector, the Cabibbo-Kobayashi-Maskawa (CKM) mixing matrix and the mass hierarchies for up and down type quarks, which can be given approximately by
\beqa
m_t:m_c:m_u\sim 1:\epsilon:\epsilon^3~,~~~m_b:m_s:m_d\sim 1:\epsilon:\epsilon^2~,
\eeqa
can also be explained by typical value of $\tau$ near the $i\infty$ fixed point in the modular $S_4^\pr$ model~\cite{Petcov:2026mdx,Abe:2023qmr}. 
For example, with the following choices of representations and modular weights for the three generations of $SU(2)_L$ doublets $Q_L\sim({\bf 3}, 2)$, up type singlets $(t_L^c,c_L^c,u_L^c)\sim({\bf \hat{1},\hat{1},1};3,1,2)$, down type singlets $(b_L^c,s_L^c,d_L^c)\sim({\bf \hat{1},{1},1};5,2,4)$,  the best-fit value based on modular $S_4^\pr$ model is given by $\epsilon=e^{-2\pi \Im\tau/4}$ for $\tau\approx 2.28 i$. From eq.(\ref{xizerotau:VEV}), such a stabilized $\tau$ value can be obtained by choosing $\tl{k}_N=9/2$. 

\section{\label{sec-5}Conclusions}

Stabilizing the modulus field near the infinite fixed point $\tau=i\infty$ with simple realizations is a well-motivated endeavor. However, the non-minimal coupling of scalars to gravity—whose presence inherently modifies the modulus scalar potential—has not yet been explored within the modular flavor framework. In this work, we investigate the modular flavor model and the stabilization of a single modulus field in Jordan frame supergravity, specifically focusing on a non-minimal scalar-curvature coupling of the form $\Phi(\tau,\bar{\tau})R$. Modular invariance, the positivity of the scale factor and positive definiteness of the Kahler metric constrain stringently the form of the frame function, and consequently the Kahler potential through the relation $\Phi(\tau,\bar{\tau})=-3\exp[-K(\tau,\bar{\tau})/3]$. We analyze the general properties of the scalar potentials following the Weyl scale transformation from the Jordan frame to the Einstein frame. We find that the shape of the resulting scalar potential in the Einstein frame differs significantly from that of ordinary single-modulus stabilization mechanisms. The resulting scalar potential can become stationary at the $i\infty$ fixed point, leading to a runaway-type vacuum. Such a runaway-type vacuum can be properly stabilized at typical modulus VEV with large $\Im\tau$.  Furthermore, we numerically evaluate modulus stabilization for several simplified scenarios, which can be readily generalized to multiple-modulus setups.

The frame function coupled to the Ricci scalar also incorporate terms related to the MSSM matter fields. Following the Weyl transformation from the Jordan frame to the Einstein frame, such a frame function would also contribute to the kinetic terms of the matter fields. In the small matter field value regions, the Yukawa couplings no longer need any scale factor after canonically normalizing the fermionic matter and Higgs fields. 

Finally, the $|H(\tau)|^2$ factor within the frame function could potentially be replaced by $\exp[\zeta|H(\tau)|^2]$ (or the exponential of a Maass form $\exp[-E(\tau,s)]$ for $s>1$), thereby accommodating parameter points where $H(\tau)=0$. We plan to explore these interesting possibilities in our future studies. Discussions on single modulus scenario in Jordan frame supergravity can be generalized to factoriable multiple moduli scenarios~\cite{FW-YKZ} and scenarios with Siegel modular form.

 \appendix
 \section{\label{appendix:A}Modular forms}
  The Dedekind $\eta$-function is a modular form with modular weight $1/2$ under modular transformation up to a phase. Its transformation under $\ga$ is given by
 \beqa
\eta(\ga\tau)=\epsilon_{\eta}(\ga)(c\tau+d)^{1/2} \eta(\tau)~, ~~\forall \ga\in \overline{\Ga}~,
\eeqa
where $\epsilon_{\eta}(\ga)$ depends on $\ga$ (but not on $\tau$) and can be expressed as
\beqa
\epsilon_{\eta}(\ga)&=&\exp\[\f{\pi i}{12}\omega(a,b,c,d)-i\f{\pi}{4}\],~
\eeqa
that satisfies $\epsilon_{\eta}(\ga)^{24}=1$~\cite{DHoker:2022dxx}. Here
\beqa
\omega(a,b,c,d)=\(\f{a+d}{c}+12s(-d,c)\),
\eeqa
is an integer and $s(-d,c)$ is a Dedekind sum.

The expansion of the $\eta(\tau)$ function is given by
\beqa
\eta(\tau)&=&q^{1/24}\sum\limits_{n=-\infty}^{\infty}(-1)^n q^{n(3n-1)/2},\nn\\
&=&q^{1/24}\[1+\sum\limits_{n=1}^{\infty}(-1)^n\(q^{n(3n-1)/2}+q^{n(3n+1)/2}\)\],\nn\\
&=&q^{1/24}\[1-q-q^2+q^5+q^7-q^{12}-\cdots\],~~~~~q\equiv e^{i 2\pi \tau}~.
\eeqa

The derivative of Dedekind $\eta$-function is given by
 \beqa
 \f{\eta^\pr(\tau)}{\eta(\tau)}=\f{i}{4\pi} G_2(\tau)~,
 \eeqa
 where the holomorphic Eisenstein series $G_2(\tau)$, which is a quasi-modular form
 \beqa
 G_2(\ga \tau)=(c\tau+d)^2G_2(\tau)-2\pi i c(c\tau+d)~,
 \eeqa
 is related to its weight two non-holomorphic counterpart $\hat{G}_2(\tau,\bar{\tau})$ as
\beqa
\hat{G}_2(\tau,\bar{\tau})=G_2(\tau) -\f{\pi}{\Im \tau}~.
\eeqa

The form of $\eta^{\pr\pr}(\tau)$ can also be obtained as
 \beqa
 \eta^{\pr\pr}(\tau)&=&\(\f{i}{4\pi}\eta(\tau) G_2(\tau)\)^\pr\nn\\
 &=&-\f{1}{16\pi^2}\eta(\tau) G_2(\tau)-\f{\pi^2}{72}\[E_2^2(\tau)-E_4(\tau)\]~,
 \eeqa 
with
\beqa
\f{\pa \hat{G}_2(\tau,\bar{\tau})}{\pa \tau}=\f{i\pi^3}{18}\(E_2^2(\tau)-E_4(\tau)\)-i\f{\pi}{2(\Im \tau)^2}~.
\eeqa

The Eisenstein series is defined as
\beqa
G_{k}(\tau) = \sum_{(m,n) \in \mathbb{Z}^2 \setminus \{(0,0)\}} \frac{1}{(m + n\tau)^{k}},
\eeqa
which is absolutely convergent for $k\geq3$, and vanish for odd $k$ in
view of the lattice symmetry~\cite{DHoker:2022dxx}. 

The Eisenstein series for even integer $k\geq 2$, which converges to a modular form of weight $k$ in the upper-half plane for $k\geq 4$, are defined as 
\beqa
G_k(\tau) = 2\zeta(k) + \frac{2(2\pi i)^k}{(k-1)!} \sum_{n=1}^\infty \sigma_{k-1}(n) q^n, \quad q = e^{2\pi i \tau},
\eeqa
where $\sigma_{k-1}(n) = \sum_{d|n} d^{k-1}$ is the divisor function. Note that 
\beqa 
G_2(i\infty)=2\zeta(2)=\f{\pi^2}{3}~,~G_4(i\infty)=2\zeta(4)=\f{\pi^4}{45}~,~G_6(i\infty)=2\zeta(6)=\f{2\pi^6}{945}~.
\eeqa

The normalized Eisenstein series $E_k(\tau)$ are defined from $G_k(\tau)$ by factoring out the $2\zeta(k)$ factor and and using the relation $(2\pi i)^k B_k=-2k!\zeta(k)$ between Bernoulli numbers and $\zeta(k)$ for even $k$
\beqa
E_k(\tau)\equiv \f{G_k(\tau)}{2\zeta(k)}=1-\f{2k}{B_k} \sum_{n=1}^\infty \sigma_{k-1}(n) q^n~,
\eeqa
which have $E_k(\tau)\ra 1$ as $\tau\ra i\infty$ for any even $k\geq 2$.

The Klein $j$-function is defined as a modular form of zero weight
\beqa
j(\tau)&\equiv& \f{3^65^3}{\pi^{12}}\f{G_4(\tau)^3}{\[\eta(\tau)\]^{24}}
=\[\f{72}{\pi^2 \eta^6(\tau)}\(\f{\eta^\pr(\tau)}{\eta^3(\tau)}\)^\pr\]^3,\nn\\
&\approx& 744+q^{-1}+196884 q+ 21493760 q^2 + 864299970 q^3 + {\cal O}(q^4)~,
\eeqa
with $j(i\infty)$ diverges, $j(\omega) = 0$ and $j(i) = 1728$. In particular, $j(\tau)$ takes real values only for CP-conserving values of $\tau$, i.e., when $\tau$ is on the border of the fundamental domain or when $\Re\tau = 0$. Note that $j(\tau)$ has a triple zero at $\tau= \omega$, and $j(\tau)-1728$ has a double zero at $\tau=i$~\cite{Klein:jfunction}. 

The derivative $j^\pr(\tau)$ is given by
\beqa
j^\pr(\tau)=-2\pi i\f{E_6(\tau) E_4^2(\tau)}{\[\eta(\tau)\]^{24}}~,
\eeqa
which vanishes only at these values: $j^\pr(\omega) = j^\prime(i) = 0$~\cite{Klein:jfunction}. Especially, we note that
\beqa
\left.\f{j^\pr(\tau)}{j(\tau)}\right|_{\tau=i\infty}&=&-\left.{2\pi i}\f{E_6(\tau)}{E_4(\tau)}\right|_{\tau=i\infty}=-2\pi i~,\nn\\
\left.\f{j^{\pr\pr}(\tau)}{j(\tau)}\right|_{\tau=i\infty}&=&
\left.\f{(2\pi)^2}{[E_4(\tau)]^2}\[\f{1}{6}E_2(\tau)E_6(\tau)E_4(\tau) -\f{1}{2}E_4^3(\tau)-\f{2}{3}E_6^2(\tau) \]\right|_{\tau=i\infty}\nn\\
&=&-4\pi^2~,
\eeqa 
is finite.
\section{Values of $H^\pr(\tau)$ and $H^{\pr\pr}(\tau)$ at typical points\label{appendix:B}}
 For non-vanishing values of $H(\tau)$, with the most general form of $\mathcal{P}(j(\tau))$ given by
\beqa
\mathcal{P}(j(\tau))=\sum\limits_{k=0}^N  c_k\[j(\tau)\]^k~,
\eeqa 
we have
\beqa
H'(\tau)&=& H(\tau) j'(\tau) \left[ \frac{n}{3} \frac{1}{j(\tau)} + \frac{m}{2} \frac{1}{j(\tau)-1728} + \frac{1}{\mathcal{P}(j(\tau))} \frac{\partial \mathcal{P}(j(\tau))}{\partial j(\tau)} \right].
\eeqa

The expression of ${H}^{\pr\pr}(\tau)$ can also be obtained as
\beqa
{H}^{\pr\pr}(\tau)&=&{H}(\tau) [j'(\tau)]^2 \left[ \frac{{n}}{3} \frac{1}{j(\tau)} + \frac{{m}}{2} \frac{1}{j(\tau)-1728} + \frac{1}{{\mathcal{P}}(j(\tau))} \frac{\partial {\mathcal{P}}(j(\tau))}{\partial j(\tau)} \right]^2\nn\\
&+&{H}(\tau)\left\{j^{\pr}(\tau)\left[ \frac{{n}}{3} \frac{1}{j(\tau)} + \frac{{m}}{2} \frac{1}{j(\tau)-1728} + \frac{1}{{\mathcal{P}}(j(\tau))} \frac{\partial {\mathcal{P}}(j(\tau))}{\partial j(\tau)} \right]\right\}^\pr~,
\label{WM:derivative}
\eeqa
when ${H}(\tau)\neq 0$. Both $H'(\tau)$ and ${H}^{\pr\pr}(\tau)$ can factor out the ${H}(\tau)$ factor when ${H}(\tau)\neq 0$.

It can be calculated that the values of $H^\pr(\tau)/H(\tau)$ and  $H^{\pr\pr}(\tau)/H(\tau)$ are finite at $\tau=i\infty$, which are given by
\beqa
\left.\f{H^\pr(\tau)}{H(\tau)}\right|_{\tau=i\infty}&\equiv& \left.j'(\tau) \left[ \frac{n}{3} \frac{1}{j(\tau)} + \frac{m}{2} \frac{1}{j(\tau)-1728} + \frac{1}{\mathcal{P}(j(\tau))} \frac{\partial \mathcal{P}(j(\tau))}{\partial j(\tau)} \right]\right|_{\tau=i\infty},\nn\\
&\sim& j'(\tau) \left[ \frac{n}{3} \frac{1}{j(\tau)} + \frac{m}{2} \frac{1}{j(\tau)-1728} + \frac{c_N N \[j(\tau)\]^{N-1}}{c_N\[j(\tau)\]^N}  \right],\nn\\
&=&-2\pi i \[\frac{n}{3}+\frac{m}{2}+N\]\equiv \tl{A}_H~,
\eeqa
while
\beqa
\left.\f{{H}^{\pr\pr}(\tau)}{H(\tau)}\right|_{\tau=i\infty}&\sim& [j^\pr(\tau)]^2 \left[ \frac{n}{3} \frac{1}{j(\tau)} + \frac{m}{2} \frac{1}{j(\tau)-1728} + \frac{N}{j(\tau)} \right]^2~\nn\\
&+&j^{\pr\pr}(\tau)\left[ \frac{n}{3} \frac{1}{j(\tau)} + \frac{{m}}{2} \frac{1}{j(\tau)-1728} + \frac{N}{j(\tau)} \right]~\nn\\
&-&[j^{\pr}(\tau)]^2\left[ \frac{n}{3} \frac{1}{j^2(\tau)} + \frac{{m}}{2} \frac{1}{\[j(\tau)-1728\]^2}+ \frac{N}{j^2(\tau)} \right],\nn\\
&=&-4\pi^2\[\frac{n}{3}+\frac{m}{2}+N\]\equiv \tl{B}_H~.
\eeqa
 The values of ${H}^\pr(\tau)$ and ${H}^{\pr\pr}(\tau)$ at the fixed point $\tau=i,\omega$ are presented here for later convenience, which had been discussed in our previous works~\cite{natural:mu}
\beqa
{H}^\pr(i)&=&\left\{\bea{c}-i\pi  (12)^{n}\f{[E_4(i)]^2}{[\eta(i)]^{12}}P(1728) , ~~~m=1\\ 0,~~~~~~~~~~~~~~~~\quad~\quad\quad\quad\quad~~m\neq 1\eea\right.\nn\\
{H}^\pr(\omega)&=&\left\{\bea{c}-\f{i 2\pi}{3} (-12)^{3m/2}\f{E_6(\omega)}{[\eta(\omega)]^{8}}P(0),~~~n=1\\0 ,~~~~~~~~~~~~~~~~~\quad\quad\quad\quad\quad~~n\neq 1\eea\right.~.
\label{Hprime}
\eeqa
and
 \beqa
 {H}^{\pr\pr}(i)&=&\left\{\bea{c}-\f{2\pi^2}{[\eta(i)]^{24}}[E_4(i)]^4\[\f{n}{3}P(1728)(12)^{n-3}+(12)^{n} P^\pr(1728)\]~,~~~~m=0\\
 ~0,~~~~~~~~~~~~~~~~~~~~~~~~~~~~~~~~~~~~~~~~~~~~~~~~~~~~~~~~~~~~~~~~~~~~~~~m=1\\
 -\f{2\pi^2}{[\eta(i)]^{24}}[E_4(i)]^4 (12)^n P(1728)~,~~~~~~~~~~~~~~~~~~~~~~~~~~~~~~~~~~~~~m=2\\
 ~0,~~~~~~~~~~~~~~~~~~~~~~~~~~~~~~~~~~~~~~~~~~~~~~~~~~~~~~~~~~~~~~~~~~~~~~~~~~m>2\eea\right.\nn\\
 {H}^{\pr\pr}(\omega)&=&\left\{\bea{c}0~,~~~~~~~~~~~~~~~~~~~~~~~~~~~~~~~~~~~~~~~~~~~~~~~~~~~~~~~~~~~~~~~~~~~n\neq 2 \\
 -\f{8\pi^2}{9}\f{[E_6(\omega)]^2}{[\eta(\omega)]^{16}}(-12)^{\f{m}{2}}P(0)~,~~~~~~~~~~~~~~~~~~~~~~~~~~~~~~~~~~~~~~n=2
 \eea\right.
 \label{Hprimeprime}
 \eeqa

\begin{acknowledgments}
This work was supported by the National Natural Science Foundation of China (NNSFC) under grant Nos.12075213 and 12335005, by the Natural Science Foundation for Distinguished Young Scholars of Henan Province under grant number 242300421046,
\end{acknowledgments}


\begin{thebibliography}{99}
\vspace{-1mm}
\bibitem{Feruglio:2017spp}
F.~Feruglio, "Are neutrino masses modular forms?",
arXiv:1706.08749 [hep-ph].


\bibitem{Feruglio:2019ybq}
F.~Feruglio and A.~Romanino,
Rev. Mod. Phys. \textbf{93} (2021) no.1, 015007
doi:10.1103/RevModPhys.93.015007
[arXiv:1912.06028 [hep-ph]].

\bibitem{Kobayashi:2023zzc}
T.~Kobayashi and M.~Tanimoto,
[arXiv:2307.03384 [hep-ph]].

\bibitem{Ding:2023htn}
G.~J.~Ding and S.~F.~King,
Rept. Prog. Phys. \textbf{87} (2024) no.8, 084201
arXiv:2311.09282 [hep-ph].

\bibitem{Ding:2024ozt}
G.~J.~Ding and J.~W.~F.~Valle,
[arXiv:2402.16963 [hep-ph]].


\bibitem{Kobayashi:2018vbk}
T.~Kobayashi, K.~Tanaka and T.~H.~Tatsuishi,
Phys. Rev. D \textbf{98} (2018) no.1, 016004
arXiv:1803.10391 [hep-ph].

\bibitem{Okada:2019xqk}
H.~Okada and Y.~Orikasa,
Phys. Rev. D \textbf{100} (2019) no.11, 115037
arXiv:1907.04716 [hep-ph].

\bibitem{Du:2020ylx}
X.~Du and F.~Wang,
JHEP \textbf{02} (2021), 221
arXiv:2012.01397 [hep-ph].







\bibitem{Petcov:2018snn}
S.~T.~Petcov and A.~V.~Titov,
Phys. Rev. D \textbf{97} (2018) no.11, 115045
arXiv:1804.00182 [hep-ph].



\bibitem{Criado:2018thu}
J.~C.~Criado and F.~Feruglio,
SciPost Phys. \textbf{5} (2018) no.5, 042
arXiv:1807.01125 [hep-ph].



\bibitem{Kobayashi:2018scp}
T.~Kobayashi, N.~Omoto, Y.~Shimizu, K.~Takagi, M.~Tanimoto and T.~H.~Tatsuishi,
JHEP \textbf{11} (2018), 196
arXiv:1808.03012 [hep-ph].


\bibitem{Okada:2018yrn}
H.~Okada and M.~Tanimoto,
Phys. Lett. B \textbf{791} (2019), 54-61
arXiv:1812.09677 [hep-ph].

\bibitem{Novichkov:2018yse}
P.~P.~Novichkov, S.~T.~Petcov and M.~Tanimoto,
Phys. Lett. B \textbf{793} (2019), 247-258
arXiv:1812.11289 [hep-ph].



\bibitem{Okada:2020ukr}
H.~Okada and M.~Tanimoto,
Phys. Rev. D \textbf{103} (2021) no.1, 015005
arXiv:2009.14242 [hep-ph].


\bibitem{Petcov:2022fjf}
S.~T.~Petcov and M.~Tanimoto,
Eur. Phys. J. C \textbf{83} (2023) no.7, 579
arXiv:2212.13336 [hep-ph].


\bibitem{CentellesChulia:2023osj}
S.~Centelles Chuli\'a, R.~Kumar, O.~Popov and R.~Srivastava,
Phys. Rev. D \textbf{109}, no.3, 035016 (2024)
doi:10.1103/PhysRevD.109.035016
[arXiv:2308.08981 [hep-ph]].

\bibitem{Kumar:2023moh}
R.~Kumar, P.~Mishra, M.~K.~Behera, R.~Mohanta and R.~Srivastava,
Phys. Lett. B \textbf{853}, 138635 (2024)
doi:10.1016/j.physletb.2024.138635
[arXiv:2310.02363 [hep-ph]].

\bibitem{Nomura:2024ctl}
T.~Nomura and H.~Okada,
arXiv:2409.10912 [hep-ph].

\bibitem{Pathak:2024sei}
G.~Pathak, P.~Das and M.~K.~Das,
arXiv:2411.13895 [hep-ph].



\bibitem{Penedo:2018nmg}
J.~T.~Penedo and S.~T.~Petcov,
Nucl. Phys. B \textbf{939} (2019), 292-307
arXiv:1806.11040 [hep-ph].
\bibitem{Novichkov:2018ovf}
P.~P.~Novichkov, J.~T.~Penedo, S.~T.~Petcov and A.~V.~Titov,
JHEP \textbf{04} (2019), 005
arXiv:1811.04933 [hep-ph].
\bibitem{Kobayashi:2019xvz}
T.~Kobayashi, Y.~Shimizu, K.~Takagi, M.~Tanimoto and T.~H.~Tatsuishi,
Phys. Rev. D \textbf{100} (2019) no.11, 115045
[erratum: Phys. Rev. D \textbf{101} (2020) no.3, 039904]
arXiv:1909.05139 [hep-ph].
\bibitem{Wang:2020dbp}
X.~Wang,
Nucl. Phys. B \textbf{962} (2021), 115247
arXiv:2007.05913 [hep-ph].

\bibitem{Qu:2021jdy}
B.~Y.~Qu, X.~G.~Liu, P.~T.~Chen and G.~J.~Ding,
Phys. Rev. D \textbf{104} (2021) no.7, 076001
arXiv:2106.11659 [hep-ph].


\bibitem{deMedeirosVarzielas:2025byb}
I.~de Medeiros Varzielas, M.~S.~Liu, A.~Sengupta and J.~Talbert,
[arXiv:2512.19789 [hep-ph]].


\bibitem{Novichkov:2018nkm}
P.~P.~Novichkov, J.~T.~Penedo, S.~T.~Petcov and A.~V.~Titov,
JHEP \textbf{04} (2019), 174
arXiv:1812.02158 [hep-ph].



\bibitem{Ding:2019xna}
G.~J.~Ding, S.~F.~King and X.~G.~Liu,
Phys. Rev. D \textbf{100} (2019) no.11, 115005
arXiv:1903.12588 [hep-ph].


\bibitem{Yao:2020zml}
C.~Y.~Yao, X.~G.~Liu and G.~J.~Ding,
Phys. Rev. D \textbf{103} (2021) no.9, 095013
arXiv:2011.03501 [hep-ph].



\bibitem{deMedeirosVarzielas:2022ihu}
I.~de Medeiros Varzielas and J.~Louren\c{c}o,
Nucl. Phys. B \textbf{984} (2022), 115974
arXiv:2206.14869 [hep-ph].


\bibitem{Abbas:2024bbv}
M.~A.~Abbas,
LHEP \textbf{2024} (2024), 545.





\bibitem{deAnda:2018ecu}
F.~J.~de Anda, S.~F.~King and E.~Perdomo,
Phys. Rev. D \textbf{101} (2020) no.1, 015028
arXiv:1812.05620 [hep-ph].

\bibitem{Kobayashi:2019rzp}
T.~Kobayashi, Y.~Shimizu, K.~Takagi, M.~Tanimoto and T.~H.~Tatsuishi,
PTEP \textbf{2020} (2020) no.5, 053B05
arXiv:1906.10341 [hep-ph].



\bibitem{Chen:2021zty}
P.~Chen, G.~J.~Ding and S.~F.~King,
JHEP \textbf{04} (2021), 239
arXiv:2101.12724 [hep-ph].

\bibitem{Zhao:2021jxg}
Y.~Zhao and H.~H.~Zhang,
JHEP \textbf{03} (2021), 002
arXiv:2101.02266 [hep-ph].

\bibitem{King:2021fhl}
S.~F.~King and Y.~L.~Zhou,
JHEP \textbf{04} (2021), 291
arXiv:2103.02633 [hep-ph].

\bibitem{Ding:2021zbg}
G.~J.~Ding, S.~F.~King and C.~Y.~Yao,
Phys. Rev. D \textbf{104} (2021) no.5, 055034
arXiv:2103.16311 [hep-ph].




\bibitem{Charalampous:2021gmf}
G.~Charalampous, S.~F.~King, G.~K.~Leontaris and Y.~L.~Zhou,
Phys. Rev. D \textbf{104} (2021) no.11, 115015
arXiv:2109.11379 [hep-ph].

\bibitem{Du:2022lij}
X.~K.~Du and F.~Wang,
JHEP \textbf{01} (2023), 036
arXiv:2209.08796 [hep-ph].


\bibitem{King:2024gsd}
S.~F.~King, G.~K.~Leontaris, L.~Marsili and Y.~L.~Zhou,
arXiv:2407.02701 [hep-ph].


\bibitem{Ding:2021eva}
G.~J.~Ding, S.~F.~King and J.~N.~Lu,
JHEP \textbf{11} (2021), 007
arXiv:2108.09655 [hep-ph].

\bibitem{Ding:2022bzs}
G.~J.~Ding, S.~F.~King, J.~N.~Lu and B.~Y.~Qu,
JHEP \textbf{10} (2022), 071
arXiv:2206.14675 [hep-ph].






\bibitem{Kobayashi:2020hoc}
T.~Kobayashi and H.~Otsuka,
Phys. Rev. D \textbf{101}, no.10, 106017 (2020)
doi:10.1103/PhysRevD.101.106017
[arXiv:2001.07972 [hep-th]].

\bibitem{Kobayashi:2020uaj}
T.~Kobayashi and H.~Otsuka,
Phys. Rev. D \textbf{102}, no.2, 026004 (2020)
doi:10.1103/PhysRevD.102.026004
[arXiv:2004.04518 [hep-th]].


\bibitem{Ishiguro:2020tmo}
K.~Ishiguro, T.~Kobayashi and H.~Otsuka,
JHEP \textbf{03}, 161 (2021)
doi:10.1007/JHEP03(2021)161
[arXiv:2011.09154 [hep-ph]].


\bibitem{NPP:2201.02020} P.P. Novichkov, J.T. Penedo and S.T. Petcov, Modular flavour symmetries and modulus stabilisation, JHEP 03 (2022) 149 [arXiv: 2201.02020].

\bibitem{King:2023snq}
S.~F.~King and X.~Wang,
Phys. Rev. D \textbf{110} (2024) no.7, 076026
doi:10.1103/PhysRevD.110.076026
[arXiv:2310.10369 [hep-ph]].


\bibitem{King:2024ssx}
S.~F.~King and X.~Wang,
JCAP \textbf{07} (2024), 073
doi:10.1088/1475-7516/2024/07/073
[arXiv:2405.08924 [hep-ph]].



\bibitem{Ding:2024neh}
G.~J.~Ding, S.~Y.~Jiang and W.~Zhao,
JCAP \textbf{10} (2024), 016
doi:10.1088/1475-7516/2024/10/016
[arXiv:2405.06497 [hep-ph]].


\bibitem{Higaki:2024pql}
T.~Higaki, T.~Kobayashi, K.~Nasu and H.~Otsuka,
JHEP \textbf{09}, 024 (2024)
doi:10.1007/JHEP09(2024)024
[arXiv:2405.18813 [hep-ph]].


\bibitem{Higaki:2024jdk}
T.~Higaki, J.~Kawamura and T.~Kobayashi,
JHEP \textbf{04} (2024), 147
doi:10.1007/JHEP04(2024)147
[arXiv:2402.02071 [hep-ph]].

\bibitem{Kobayashi:2023spx}
T.~Kobayashi, K.~Nasu, R.~Sakuma and Y.~Yamada,
Phys. Rev. D \textbf{108} (2023) no.11, 115038
doi:10.1103/PhysRevD.108.115038
[arXiv:2310.15604 [hep-ph]].

\bibitem{Funakoshi:2024yxg}
S.~Funakoshi, J.~Kawamura, T.~Kobayashi, K.~Nasu and H.~Otsuka,
[arXiv:2409.19261 [hep-th]].



\bibitem{natural:mu} Hong-jie Fan, Fei Wang, Ying Kai Zhang, Natural solution of SUSY $\mu$ problem from modulus stabilization in modular flavor model, Phys. Rev. D \textbf{112}, 115040 (2025) 
doi: https://doi.org/10.1103/8z51-3zgf [arXiv:2412.07642[hep-ph]].



\bibitem{target:space} M. Cvetic, A. Font, L.E. Ibanez, D. Lust, F. Quevedo, Nucl.Phys. B
361(1991)194-232.

\bibitem{PLNSR:2304.14437} V. Knapp-Perez, X.-G. Liu, H.P. Nilles, S. Ramos-Sanchez and M. Ratz, Matter matters in moduli fixing and modular flavor symmetries, [arXiv:2304.14437].


\bibitem{Higaki:2024ueb}
T.~Higaki, J.~Kawamura, T.~Kobayashi, K.~Nasu and R.~Sakuma,
JHEP \textbf{05} (2025), 111
doi:10.1007/JHEP05(2025)111
[arXiv:2412.18435 [hep-ph]].

\bibitem{Leedom:2022zdm}
J.~M.~Leedom, N.~Righi and A.~Westphal,
JHEP \textbf{02} (2023), 209
doi:10.1007/JHEP02(2023)209
[arXiv:2212.03876 [hep-th]].

\bibitem{Gonzalo:2018guu}
E.~Gonzalo, L.~E.~Ib\'a\~nez and \'A.~M.~Uranga,
JHEP \textbf{05} (2019), 105
doi:10.1007/JHEP05(2019)105
[arXiv:1812.06520 [hep-th]].




\bibitem{Baur:2019kwi}
A.~Baur, H.~P.~Nilles, A.~Trautner and P.~K.~S.~Vaudrevange,
Phys. Lett. B \textbf{795} (2019), 7-14
doi:10.1016/j.physletb.2019.03.066
[arXiv:1901.03251 [hep-th]].


\bibitem{Novichkov:2019sqv}
P.~P.~Novichkov, J.~T.~Penedo, S.~T.~Petcov and A.~V.~Titov,
JHEP \textbf{07} (2019), 165
doi:10.1007/JHEP07(2019)165
[arXiv:1905.11970 [hep-ph]].

\bibitem{Ishiguro:2020nuf}
K.~Ishiguro, T.~Kobayashi and H.~Otsuka,
Nucl. Phys. B \textbf{973} (2021), 115598
doi:10.1016/j.nuclphysb.2021.115598
[arXiv:2010.10782 [hep-th]].

\bibitem{Feruglio:2021dte}
F.~Feruglio, V.~Gherardi, A.~Romanino and A.~Titov,
JHEP \textbf{05} (2021), 242
doi:10.1007/JHEP05(2021)242
[arXiv:2101.08718 [hep-ph]].











\bibitem{Futamase:1987ua}
  T.~Futamase and K.~I.~Maeda,
 \textit{Chaotic inflationary scenario in models having nonminimal coupling with
  curvature},
  Phys.\ Rev.\  \textbf{D39}, 399 (1989).

\bibitem{Salopek:1988qh}
  D.~S.~Salopek, J.~R.~Bond and J.~M.~Bardeen,
\textit{Designing density fluctuation spectra in inflation},
  Phys.\ Rev.\  \textbf{D40}, 1753 (1989).

\bibitem{Makino:1991sg}
  N.~Makino and M.~Sasaki,
  \textit{The Density perturbation in the chaotic inflation with nonminimal
coupling},
  Prog.\ Theor.\ Phys.\  \textbf{86}, 103 (1991).

\bibitem{Fakir:1992cg}
  R.~Fakir, S.~Habib and W.~Unruh,
  \textit{Cosmological density perturbations with modified gravity},
  Astrophys.\ J.\  \textbf{394}, 396 (1992).







\bibitem{Higgsinflation:A}  F.~L.~Bezrukov and M.~Shaposhnikov, ``The Standard
Model Higgs boson as the inflaton,'' Phys.\ Lett.\ B {\bf 659}, 703
(2008) [arXiv:0710.3755 [hep-th]].


\bibitem{Bezrukov:2008ut}
  F.~Bezrukov, D.~Gorbunov and M.~Shaposhnikov,
  ``On initial conditions for the Hot Big Bang,''
  JCAP {\bf 0906}, 029 (2009) [arXiv:0812.3622 [hep-ph]].
 
\bibitem{GarciaBellido:2008ab}
  J.~Garcia-Bellido, D.~G.~Figueroa and J.~Rubio,
  ``Preheating in the Standard Model with the Higgs-Inflaton coupled to
  gravity,'' Phys.\ Rev.\ D {\bf 79}, 063531 (2009) [arXiv:0812.4624 [hep-ph]].
  
\bibitem{DeSimone:2008ei}
  A.~De Simone, M.~P.~Hertzberg and F.~Wilczek,
``Running Inflation in the Standard Model,''
  Phys.\ Lett.\  B {\bf 678}, 1 (2009) [arXiv:0812.4946 [hep-ph]].
  
\bibitem{Bezrukov:2009db}
  F.~Bezrukov and M.~Shaposhnikov,
``Standard Model Higgs boson mass from inflation: two loop
analysis,''
  JHEP {\bf 0907}, 089 (2009) [arXiv:0904.1537 [hep-ph]].
  
  
\bibitem{Barvinsky:2009fy}
  A.~O.~Barvinsky, A.~Y.~Kamenshchik, C.~Kiefer, A.~A.~Starobinsky and C.~Steinwachs,
  ``Asymptotic freedom in inflationary cosmology with a non-minimally coupled
  Higgs field,'' JCAP {\bf 0912}, 003 (2009) [arXiv:0904.1698 [hep-ph]].


\bibitem{Einhorn:2009bh}
M.~B.~Einhorn and D.~R.~T.~Jones,
JHEP \textbf{03} (2010), 026
doi:10.1007/JHEP03(2010)026
[arXiv:0912.2718 [hep-ph]].





\bibitem{Ferrara:2010yw}
S.~Ferrara, R.~Kallosh, A.~Linde, A.~Marrani and A.~Van Proeyen,
Phys. Rev. D \textbf{82} (2010), 045003
doi:10.1103/PhysRevD.82.045003
[arXiv:1004.0712 [hep-th]].

\bibitem{Lee:2010hj}
H.~M.~Lee,
JCAP \textbf{08} (2010), 003
doi:10.1088/1475-7516/2010/08/003
[arXiv:1005.2735 [hep-ph]].



\bibitem{Kallosh:2000ve}
R.~Kallosh, L.~Kofman, A.~D.~Linde and A.~Van Proeyen,
Class. Quant. Grav. \textbf{17} (2000), 4269-4338
[erratum: Class. Quant. Grav. \textbf{21} (2004), 5017]
doi:10.1088/0264-9381/17/20/308
[arXiv:hep-th/0006179 [hep-th]].

\bibitem{Ferrara:2010in}
S.~Ferrara, R.~Kallosh, A.~Linde, A.~Marrani and A.~Van Proeyen,
Phys. Rev. D \textbf{83} (2011), 025008
doi:10.1103/PhysRevD.83.025008
[arXiv:1008.2942 [hep-th]].

\bibitem{Cremmer:1978hn} E.~Cremmer, B.~Julia, J.~Scherk, S.~Ferrara,
L.~Girardello and P.~van Nieuwenhuizen, ``Spontaneous Symmetry
Breaking And Higgs Effect In Supergravity Without Cosmological
Constant,'' Nucl.\ Phys.\
 B {\bf 147}, 105 (1979).

\bibitem{BFNS-82} R.~Barbieri, S.~Ferrara, D.~V.~Nanopoulos and K.~S.~Stelle,
``Supergravity, R Invariance And Spontaneous Supersymmetry Breaking,''
  Phys.\ Lett.\  B {\bf 113}, 219 (1982).

\bibitem{CFGVP-1}  E.~Cremmer, S.~Ferrara, L.~Girardello and A.~Van
Proeyen, ``Yang-Mills Theories With Local Supersymmetry:
Lagrangian, Transformation Laws And Superhiggs Effect,'' Nucl.\ Phys.\
B {\bf 212}, 413 (1983).

\bibitem{Girardi:1984eq}
  G.~Girardi, R.~Grimm, M.~Muller and J.~Wess,
  ``Superspace Geometry And The Minimal, Nonminimal, And New Minimal
 Supergravity Multiplets,''
  Z.\ Phys.\   C {\bf 26}, 123 (1984).

\bibitem{Wess:1992cp}
  J.~Wess and J.~Bagger,
  \textit{Supersymmetry and supergravity},
(Princeton University Press, Princeton, 1992).

\bibitem{Chen:2019ewa}
M.~C.~Chen, S.~Ramos-S{\'a}nchez and M.~Ratz,
Phys. Lett. B \textbf{801} (2020), 135153
doi:10.1016/j.physletb.2019.135153
[arXiv:1909.06910 [hep-ph]].

\bibitem{Baur:2019kwi}
A.~Baur, H.~P.~Nilles, A.~Trautner and P.~K.~S.~Vaudrevange,
Phys. Lett. B \textbf{795} (2019), 7-14
doi:10.1016/j.physletb.2019.03.066
[arXiv:1901.03251 [hep-th]].


\bibitem{FW-YKZ} Fei Wang, Ying Kai Zhang, Modulus Stabilization in Factoriable Mulitple Modular Flavor Models, to appear soon. 

\bibitem{Garcia-Bellido:2008ycs}
J.~Garcia-Bellido, D.~G.~Figueroa and J.~Rubio,
Phys. Rev. D \textbf{79} (2009), 063531
doi:10.1103/PhysRevD.79.063531
[arXiv:0812.4624 [hep-ph]].

\bibitem{Rubio:2018ogq}
J.~Rubio,
Front. Astron. Space Sci. \textbf{5} (2019), 50
doi:10.3389/fspas.2018.00050
[arXiv:1807.02376 [hep-ph]].

\bibitem{Kachru:2003aw}
S.~Kachru, R.~Kallosh, A.~D.~Linde and S.~P.~Trivedi,
Phys. Rev. D \textbf{68} (2003), 046005
doi:10.1103/PhysRevD.68.046005
[arXiv:hep-th/0301240 [hep-th]].


\bibitem{Novichkov:2021evw}
P.~P.~Novichkov, J.~T.~Penedo and S.~T.~Petcov,
JHEP \textbf{04} (2021), 206
doi:10.1007/JHEP04(2021)206
[arXiv:2102.07488 [hep-ph]].


\bibitem{Petcov:2023vws}
S.~T.~Petcov and M.~Tanimoto,
JHEP \textbf{08} (2023), 086

\bibitem{Petcov:2026mdx}
S.~T.~Petcov and M.~Tanimoto,
[arXiv:2601.04529 [hep-ph]].


\bibitem{Abe:2023qmr}
Y.~Abe, T.~Higaki, J.~Kawamura and T.~Kobayashi,
Phys. Lett. B \textbf{842} (2023), 137977
doi:10.1016/j.physletb.2023.137977
[arXiv:2302.11183 [hep-ph]].

\bibitem{DHoker:2022dxx}
E.~D'Hoker and J.~Kaidi,
[arXiv:2208.07242 [hep-th]].


\bibitem{Klein:jfunction} T. M. Apostol, Modular Functions and Dirichlet Series in Number Theory,
Graduate Texts in Mathematics. Springer, New York, NY, 1990.

\end{thebibliography}
\end{document}